\documentclass[aps,prd,preprint,superscriptaddress]{revtex4-1}
\usepackage{amsmath,amssymb}
\usepackage[colorlinks=true,urlcolor=blue,citecolor=blue,linkcolor=blue]{hyperref}

\begin{document}

\title{Analogue simulation of gravitational waves in a $3+1$ dimensional Bose-Einstein condensate}

\author{Daniel Hartley}
\email[Corresponding author: ]{daniel.hartley@univie.ac.at}
\author{Tupac Bravo}
\author{Dennis R{\"a}tzel}
\author{Richard Howl}
\affiliation{Faculty of Physics, University of Vienna, Boltzmanngasse 5, 1090 Wien, Austria}
\author{Ivette Fuentes}
\affiliation{Faculty of Physics, University of Vienna, Boltzmanngasse 5, 1090 Wien, Austria}
\affiliation{School of Mathematical Sciences, University of Nottingham, University Park, Nottingham NG7 2RD, United Kingdom}

\date{\today}

\begin{abstract}
The recent detections of gravitational waves (GWs) by the LIGO and
Virgo collaborations have opened the field of GW astronomy, intensifying
interest in GWs and other possible detectors sensitive in different
frequency ranges. Although strong GW producing events are rare and
currently unpredictable, GWs can in principle be simulated in analogue
systems at will in the lab. Simulation of GWs in a manifestly quantum
system would allow for the study of the interaction of quantum phenomena
with GWs. Such predicted interaction is exploited in a recently proposed
Bose-Einstein condensate (BEC) based GW detector. In this paper, we
show how to manipulate a BEC to mimic the effect of a passing GW.
By simultaneously varying the external potential applied to the BEC,
and an external magnetic field near a Feshbach resonance, we show
that the resulting change in speed of sound can directly reproduce
a GW metric. We also show how to simulate a metric used in the recently
proposed BEC based GW detector, to provide an environment for testing
the proposed metrology scheme of the detector. Explicit expressions
for simulations of various GW sources are given. This result is also
useful to generally test the interaction of quantum phenomena with
GWs in a curved spacetime analogue experiment.
\end{abstract}


\maketitle

\section{Introduction}

In the 36 years since the seminal proposal of Unruh to measure an
acoustic analogue to Hawking radiation from a ``sonic horizon''
in a fluid \cite{Unruh1981}, interest in analogue simulation of gravitational
fields has grown from theoretical proposals to experiments in numerous
systems. These include Bose-Einstein condensates (BECs) \cite{Lahav2010,Hung2013,Steinhauer2014},
water waves \cite{Weinfurtner2011,Euve2015} and optical fibres \cite{Philbin2008}
among others. Particular interest has been shown in using the phonon
field in a BEC, since this is a quantum system and so allows for the
study of how non-classical properties, such as entanglement, are modified
or generated by simulated gravitational fields. This has recently
culminated in the first observation of the entanglement of acoustic
Hawking radiation \cite{Steinhauer2016}, potentially providing clues
to fundamental questions for quantum gravity, such as the information
paradox. In addition to Hawking radiation from a waterfall horizon,
other proposed simulations using BECs have included conformal Schwarzschild
black holes \cite{Cropp2016,Dey2016}, rotating black holes \cite{Giacomelli2017},
FRW geometries \cite{Barcelo2003,Fagnocchi2010}, inflation \cite{Fischer2004,Cha2017}
and extensions to Einstein's general relativity, such as aether fields
\cite{Cropp2016}. In past work \cite{Bravo2015}, two of us have
considered the simulation of gravitational waves (GWs) in $1+1$ dimensions.
These are perturbations of spacetime generated by a changing quadrupole
moment of a mass distribution, and have recently been detected in
a milestone moment in science \cite{Abbott2016,Abbott2016a,Abbott2017,Abbott2017a,Abbott2017b}.
This has led to a new field of GW astronomy, enabling the exploration
of the Universe through gravitational as well as electromagnetic radiation.
Simulating GWs in fluid systems could be of astronomical interest,
for example, in studying GWs in numerically challenging, strong-field
regimes. Furthermore, since BECs are quantum systems, this could enable
the study of predicted effects such as particle creation in GW backgrounds
\cite{Jones2017}, and quantum decoherence due to GWs \cite{Jaekel2006}.
While we are interested in simulating the effect of GWs, simulating
the evolution of a GW itself on a curved background has also been proposed in
\cite{Fernandez-Corbaton2015}, where a metamaterial emulates a curved background
space-time and two-photon states model the evolution of a GW on that background.

Simulating GWs with BECs could also be used in studies of a proposed
BEC GW detector \cite{Sabin2014,Sabin2016,Howl2016a}. This detector
consists of a BEC constrained to a rigid trap with a prepared quantum
state of phonons, such as a two-mode squeezed state. The transformation
induced by the GW produces mode-mixing and phonon creation in a
phenomenon resembling the dynamical Casimir effect \cite{Moore1970,Fulling1976},
making the final state distinguishable from the probe state, i.e. decreasing
the fidelity between the initial and final states. The lower the fidelity between
the probe and final state, the better the estimation. Non-classical squeezed states
allow for quantum metrology techniques resulting in better estimation than
a classical device. At resonance, mode-mixing or phonon creation are maximised
giving rise to optimal parameter estimation.
Such a quantum resonance process is absent in laser interferometers
since the frequencies of the GWs are far from the optical regime.
However, using resonance to detect GWs was the concept behind the
first GW detector proposals, Weber bars, which are metal objects measuring
metres in length. The GW resonance in the BEC detector is similar
since the much smaller size, $\mathcal{O}\left(\mu m\right)$, is
compensated for by a much smaller speed of sound, $\mathcal{O}\left(mm/s\right)$
compared to $\mathcal{O}\left(km/s\right)$ \cite{Howl2016}. However,
the BEC detector can be cooled to considerably lower temperatures,
$\mathcal{O}\left(nK\right)$, and is a strictly quantum device. This
allows for the use of quantum metrology and, therefore, sensitivities
that are inaccessible to classical devices \cite{Braunstein1994}. A discussion
of the viability of such a detector and further details can be found in \cite{Sabin2014}.
Further studies of the viability of such a detector are in progress and this
article is part of this effort. Simulating the effect of a GW derived in \cite{Sabin2014}
could be useful for testing the metrological scheme proposed in \cite{Sabin2014,Sabin2016}.

Here we extend the work on $1+1$ GWs to the simulation of $3+1$
GWs in BECs in a covariant formalism \cite{Bravo2015}. With this
extension, all properties of a GW can be simulated, such as its polarization
and propagation vector, and the conformal factor in front of the analogue
metric no longer diverges. The paper is outlined as follows: in Section
\ref{sec:metrics} we present the spacetime metric of a GW and the acoustic
metric in a BEC. In Section \ref{sec:results} we derive and demonstrate the simulation of a GW
metric in $3+1$ dimensions, as well as the metric derived in \cite{Sabin2014}.
Examples of GWs are presented in Section \ref{sec:examples}, giving
explicit forms of flow velocities needed to simulate the effect of
commonly investigated GW sources, including compact binary inspirals
and neutron star spin down. Section \ref{sec:1+1} reduces the metric
derived in Section \ref{sec:results} to $1+1$ dimensions, and compares
this to previously published work in \cite{Bravo2015}, and we conclude
in Section \ref{sec:conclusion}.

\subsection{Definitions and conventions \label{subsec:def}}

Throughout this paper, we use the metric signature $\left(-,+,+,+\right)$,
the coordinates used are Minkowski coordinates given by $\left(ct,x,y,z\right)$
unless otherwise stated, and the Minkowski metric in these coordinates
is given by
\begin{equation}
\eta_{\mu\nu}=diag\left(-1,1,1,1\right).\label{eq:minkowski}
\end{equation}

\section{Gravitational waves and the acoustic metric \label{sec:metrics}}

\subsection{GW spacetime metric \label{subsec:gwdef}}

We first consider the general form of the metric tensor perturbed
by a GW. For a single source GW far from the source, the metric tensor
can be expressed as \cite{MTW,MaggioreGW}
\begin{equation}
g_{\mu\nu}^{\left(gw\right)}=\eta_{\mu\nu}+\epsilon h_{\mu\nu},\label{eq:gwmetric-1}
\end{equation}
where $\eta_{\mu\nu}$ is the flat Minkowski metric defined in Section
\ref{subsec:def}, and $h_{\mu\nu}$ is some perturbation corresponding
to the passing GW, parameterised by $\epsilon$, where $\left|\epsilon\right|\ll1$.
Standard notation omits this $\epsilon$ and applies the condition
$\left|h_{\mu\nu}\right|\ll1$, but we use $\epsilon$ here as a global
perturbation scale factor for consistency and clarity.

\subsubsection{Transverse traceless gauge}

This perturbation $h_{\mu\nu}$ can be expressed in the transverse
traceless (TT) gauge, in coordinates $x_{TT}^{\mu}$, for a GW travelling
in the $\hat{z}$ direction as \cite{MaggioreGW}
\begin{equation}
h_{\mu\nu}^{TT}=\begin{pmatrix} 0 & 0 & 0 & 0\\
0 & h_{+}\left(t\right) & h_{\times}\left(t\right) & 0\\
0 & h_{\times}\left(t\right) & -h_{+}\left(t\right) & 0\\
0 & 0 & 0 & 0
\end{pmatrix},\label{eq:ttgw}
\end{equation}
where $h_{+}$ and $h_{\times}$ are time-dependent functions corresponding
to two ``polarisations'' of the GW. These $h_{+}$ and $h_{\times}$
functions are typically called ``strain'' functions. We ignore the
$z$ dependence of the strain functions, as the wavelength of a GW
is typically much longer than the width of a BEC (for example, the
GWs detected by the LIGO and Virgo collaborations have wavelengths
exceeding $10^{6}$ m). Outside the source of the GWs, these strain
functions obey the simple wave equation
\begin{equation}
\eta^{\rho\sigma}\partial_{\rho}\partial_{\sigma}h_{\mu\nu}^{TT}=0.
\end{equation}
We introduce the GW metric in this gauge, as it is the clearest and
most widely known, despite not necessarily being the most physically
useful.

\subsubsection{Fermi normal coordinates \label{subsec:pdframe}}

Any metric in linearised gravity of the form of Eq. (\ref{eq:gwmetric-1})
has a ``gauge freedom'', namely a choice of small coordinate transformation
in any arbitrary direction. Consider a linearised coordinate transformation
with some function $\zeta$, such that
\begin{equation}
x^{\mu}\rightarrow x^{\mu}+\epsilon\zeta^{\mu}\left(x^{\mu}\right).\label{eq:xi1-1}
\end{equation}
Under such a coordinate transformation, the metric in Eq. (\ref{eq:gwmetric-1})
transforms as
\begin{equation}
\eta_{\mu\nu}+\epsilon h_{\mu\nu}\rightarrow\eta_{\mu\nu}+\epsilon\left[h_{\mu\nu}-\partial_{\nu}\zeta_{\mu}-\partial_{\mu}\zeta_{\nu}\right]+\mathcal{O}\left(\epsilon^{2}\right).
\end{equation}
Hence, making a coordinate transformation $x_{TT}^{\mu}\rightarrow x{}^{\mu}=x_{TT}^{\mu}+\epsilon\zeta^{\mu}$
from the TT gauge with the function
\begin{equation}
\begin{split}
\zeta^{\mu}=&\left(\frac{1}{4c}\left(2xy\partial_{t}h_{\times}+\left(x^{2}-y^{2}\right)\partial_{t}h_{+}\right),\right.\\
&\hphantom{((}\left.\frac{1}{2}\left(xh_{+}+yh_{\times}\right),\frac{1}{2}\left(xh_{\times}-yh_{+}\right),0\right),
\end{split}
\end{equation}
the metric perturbation in Eq. (\ref{eq:ttgw}) in these new coordinates
is
\begin{equation}
h_{\mu\nu}^{TT}\rightarrow h_{\mu\nu}=\begin{pmatrix}-h_{00} & \boldsymbol{0}^{\text{T}}\\
\boldsymbol{0} & \mathbb{I}_{3}
\end{pmatrix}+\mathcal{O}\left(\epsilon^{2}\right)\label{eq:gwmetric-2}
\end{equation}
where $\mathbb{I}_{n}$ is the $n$ dimensional identity matrix, and
\begin{equation}
h_{00}=-\frac{1}{2c^{2}}\left(2xy\partial_{t}^{2}h_{\times}+\left(x^{2}-y^{2}\right)\partial_{t}^{2}h_{+}\right).
\end{equation}
These coordinates are Fermi normal coordinates, and are the inertial
frame limit of the ``proper detector frame''. This metric perturbation
can also be derived by considering coordinates matching proper length
and time, then linearising the metric with respect to $R=\sqrt{\eta_{ij}x^{i}x^{j}}$
(derived for example in \cite{MaggioreGW} Section 1.3.3). One of
the most useful features of these coordinates is that they match the
laboratory coordinates of an experiment in free fall, e.g. a drag
free satellite orbiting the Earth. It is also a good approximation
for the suspended mirrors of the LIGO experiments. For notational
convenience, we also define
\begin{equation}
\begin{split}H_{00}\left(t,x,y\right) & =\int_{0}^{t}h_{00}\left(t',x,y\right)cdt'\\
 & =-\frac{1}{c}\left(xy\partial_{t}h_{\times}+\frac{1}{2}\left(x^{2}-y^{2}\right)\partial_{t}h_{+}\right).
\end{split}
\end{equation}

\subsection{Acoustic metric \label{subsec:effmetric}}

To simulate a GW metric in a BEC, we will follow the description of
a BEC on a general background metric given in \cite{Fagnocchi2010,Visser2010,Bruschi2014}.
This description models the BEC as a barotropic, irrotational
and inviscid fluid, in a covariant formalism. We are interested in
i) simulating a spacetime metric using a quantum system and ii) simulating
the effects of spacetime dynamics on a phononic field. In both cases we
require a covariant formalism that enables us to properly describe a
general relativistic spacetime and its effects on quantum fields. The
formalism developed in \cite{Fagnocchi2010,Visser2010,Bruschi2014}
enables us to do so. We point out that the system that we consider here
is a regular BEC, as those currently demonstrated in the laboratory. 
This system is usually described with non-relativistic quantum mechanics.
However, in ii) we are taking into account the underlying spacetime background
on which the BEC sits on, which requires the covariant treatment mentioned
above. We point out that we are not considering a system that is moving with
relativistic speeds or has excitations with relativistic energies. Such a
relativistic system would also need to be described by the same formalism
since covariance is also necessary.

Any BEC which can be described as a superfluid is automatically barotropic
and inviscid. In the superfluid regime, the BEC is described by a classical mean
field $\phi$ expressed as
\begin{equation}
\phi=\sqrt{\rho}e^{i\theta},
\end{equation}
with quantum fluctuations $\hat{\psi}$, defined in terms of the total
field $\hat{\Phi}$ as
\begin{equation}
\hat{\Phi}=\phi\left(1+\hat{\psi}\right).
\end{equation}
We are interested in the behaviour of these fluctuations in the ``phononic''
regime. The relativistic phononic regime condition can be written
explicitly as \cite{Fagnocchi2010}
\begin{equation}
\left|k\right|\ll\frac{\sqrt{2}}{\xi}\left(1+\frac{\hbar^{2}}{2m^{2}\xi^{2}u_{0}^{2}}\right)\text{min}\left[1,\frac{m u_0 \xi}{\sqrt{2}\hbar}\right],\label{eq:disp}
\end{equation}
where $k$ is the spatial frequency of a phononic excitation of the
BEC, $u$ is the flow velocity of the BEC defined as
\begin{equation}
u_{\mu}=\frac{\hbar}{m}\partial_{\mu}\theta\label{eq:udef}
\end{equation}
and $\xi$ is the ``healing length'' defined as
\begin{equation}
\xi=\frac{1}{\sqrt{\lambda\rho}}.\label{eq:healing}
\end{equation}
$\lambda$ encodes the strength of the interaction, defined in terms
of the interaction potential $U$ as
\begin{equation}
U\left(\phi^{\dagger}\phi,\lambda\right)=\frac{1}{2}\lambda\left|\phi^{\dagger}\phi\right|^{2}+\cdots
\end{equation}
where extra terms are $\phi^{6}$ interactions and higher, which we
ignore here. The interaction strength $\lambda$ is related to the
scattering length $a$ by
\begin{equation}
\lambda=8\pi a.
\end{equation}
Taking the non-relativistic limit of this condition, we find that
the phonons should have wavelengths far longer than the healing length
$\xi$. In this regime, and with certain additional assumptions about
the mean field properties, the fluctuations
obey a relativistic Klein-Gordon-like equation
\begin{equation}
\frac{1}{\sqrt{-G}}\partial_{\mu}\left(\sqrt{-G}G^{\mu\nu}\partial_{\nu}\hat{\psi}\right)=0\label{eq:relkg}
\end{equation}
for a tensor $G_{\mu\nu}$ with determinant $G$, called the ``acoustic
metric''. The general acoustic metric is given by (see Appendix \ref{sec:metricderivation})
\begin{equation}
G_{\mu\nu}=\frac{\rho c}{c_{s}}\left(g_{\mu\nu}+r\frac{v_{\mu}v_{\nu}}{c^{2}}\right),\label{eq:effmetric}
\end{equation}
where $g_{\mu\nu}$ is the background spacetime metric, $\rho$ is
the bulk density defined as $\rho=\phi^{*}\phi$, $r$ is related
to the speed of sound $c_{s}$ as
\begin{equation}
r=1-\frac{c_{s}^{2}}{c^{2}}
\end{equation}
and $v$ is the normalised flow velocity defined as
\begin{equation}
v_{\mu}=\frac{cu_{\mu}}{\left|u\right|}.
\end{equation}
The speed of sound $c_{s}$ is defined as
\begin{equation}
c_{s}^{2}=\frac{c^{2}c_{0}^{2}/\left|u\right|^{2}}{1+c_{0}^{2}/\left|u\right|^{2}},\label{eq:csdef}
\end{equation}
where
\begin{equation}
c_{0}^{2}=\frac{\hbar^{2}}{2m^{2}}\rho\partial_{\rho}^{2}U\left(\rho,\lambda\right)=\frac{\hbar^{2}}{2m^{2}}\lambda\rho.\label{eq:c0def}
\end{equation}
The flow velocity $v$ is normalised as
\begin{equation}
g^{\mu\nu}v_{\mu}v_{\nu}=-c^{2}.\label{eq:vnorm}
\end{equation}
Note that the definition of the flow velocity imposes irrotationality,
i.e.
\begin{equation}
\partial_{\mu}u_{\nu}=\partial_{\nu}u_{\mu}.\label{eq:irrotational}
\end{equation}
Due to the global phase symmetry of the Lagrangian that Eq. (\ref{eq:relkg})
is derived from, there is a conserved current. The conservation of
this current can be expressed as
\begin{equation}
\nabla_{\mu}\left(\rho u^{\mu}\right)=0,\label{eq:continuity}
\end{equation}
also called the continuity equation, where $\nabla_{\mu}$ is the
covariant derivative with respect to $g_{\mu\nu}$. The velocity normalisation
$\left|u\right|$ and density $\rho$ can also be directly related
to the internal and external potentials $U$ and $V$ as
\begin{equation}
\left|u\right|^{2}=c^{2}+\frac{\hbar^{2}}{m^{2}}\left\{ V+\frac{\partial U\left(\rho,\lambda\right)}{\partial\rho}-\frac{\nabla_{\mu}\nabla^{\mu}\sqrt{\rho}}{\sqrt{\rho}}\right\} .\label{eq:u=00003Dc+stuff}
\end{equation}

\section{Gravitational wave simulation \label{sec:results}}

In this paper, we present two results corresponding to different types
of simulation. The first in Section \ref{subsec:gwmetricsim} is a
direct simulation of the GW metric in Eq. (\ref{eq:gwmetric-1}).
The second result in Section \ref{subsec:gweffectsim} is a simulation
of the acoustic metric derived in \cite{Sabin2014}, to test the proposed
metrological scheme.

\subsection{GW metric simulation \label{subsec:gwmetricsim}}

The goal of this section is to directly simulate a GW spacetime, such
that the acoustic metric $G_{\mu\nu}$ has the form
\begin{equation}
G_{\mu\nu}^{\left(GW\right)}=\eta_{\mu\nu}+h_{\mu\nu}.\label{eq:directgoal}
\end{equation}
We start with the GW metric in Fermi normal coordinates, as introduced
in Section \ref{subsec:pdframe}. If we consider a BEC at rest in
these coordinates, i.e. $v_{\mu}=-c\delta_{\mu}^{0}$, where the background
metric is the flat Minkowski metric, then the acoustic metric has
the form
\begin{equation}
G_{\mu\nu}^{\left(SIM\right)}=\frac{\rho c}{c_{s}}\begin{pmatrix}-c_{s}^{2}/c^{2} & \boldsymbol{0}^{\text{T}}\\
\boldsymbol{0} & \mathbb{I}_{3}
\end{pmatrix}.\label{eq:gwmetricsim}
\end{equation}
It should be noted that the density is not completely unrestricted;
the choice of flow velocity restricts the density through the continuity
equation. From the definition of the flow velocity in Eq. (\ref{eq:udef})
and the choice of normalised velocity above,
\begin{equation}
\frac{\hbar}{m}\partial_{\mu}\theta=-\left|u\right|c\delta_{\mu}^{0}
\end{equation}
which implies that $\left|u\right|$ can only be a function of time.
Hence, with the continuity equation (Eq. (\ref{eq:continuity})),
for this particular choice of normalised velocity, we must have
\begin{equation}
\frac{\partial_{t}\rho}{\rho}=-\frac{\partial_{t}\left|u\right|}{\left|u\right|}.
\end{equation}

\subsubsection{Bulk properties}

Comparison of Eq. (\ref{eq:gwmetric-2}) and Eq. (\ref{eq:gwmetricsim})
suggests that, to simulate the GW metric, we must modulate the speed
of sound as
\begin{equation}
c_{s}^{2}=c_{s0}^{2}\left(1+\epsilon h_{00}\right),
\end{equation}
where $c_{s0}$ is the speed of sound in the absence of a simulated
GW. If we rescale the time coordinate as
\begin{equation}
c_{s0}\tau=\left(\frac{c_{s0}^{2}}{c^{2}}\right)ct,
\end{equation}
then the acoustic metric has the form
\begin{equation}
G_{\mu'\nu'}^{\left(SIM\right)}=\frac{\rho c}{c_{s}}\begin{pmatrix}-1-\epsilon h_{00} & \boldsymbol{0}^{\text{T}}\\
\boldsymbol{0} & \mathbb{I}_{3}
\end{pmatrix}\label{eq:metricsimresult}
\end{equation}
which matches the desired metric in Eq. (\ref{eq:directgoal}) up
to a conformal factor. The conformal factor will be discussed further
later in this subsection, as well as in Section \ref{subsec:conformal}.

\subsubsection{Implementation \label{subsec:implement1}}

As mentioned earlier in Section \ref{subsec:gwmetricsim}, we must
have $\left|u\right|$ dependent on time only. Furthermore, having
a density that is changing in time in the absence of flows implies
changing the local atom number density in the BEC in a uniform and
precisely controlled way, which seems experimentally unfeasible. If
it is possible to control the density independently of the flows,
then modulating the density to match the speed of sound sets the conformal
factor in Eq. (\ref{eq:metricsimresult}) to be constant, and thus
physically irrelevant. If it is not feasible to control the density
without inducing flows, then we must conclude that the density must
be constant in time, and thus
\begin{equation}
\partial_{t}\left|u\right|=-\frac{\left|u\right|\partial_{t}\rho}{\rho}=0,
\end{equation}
so $\left|u\right|$ is constant in both space and time. This conclusion
can also be drawn from the chemical potential $\mu$ of the BEC. If
the BEC is stationary, then $\mu$ is constant. Since
\begin{equation}
\mu=\frac{i\hbar\partial_{t}\phi}{\phi}=mc\left|u\right|,
\end{equation}
we can conclude that $\left|u\right|$ is constant. Using Eq. (\ref{eq:u=00003Dc+stuff}),
we can see that all of these restrictions on the density, speed of
sound and flows are only achievable by balancing the external potential
$V$ and internal interaction strength $\lambda$ to modulate the
speed of sound while leaving the density constant in time. Specifically,
we must have
\begin{equation}
V=-\lambda\rho-\frac{m^{2}}{\hbar^{2}}\left(c^{2}-\left|u\right|^{2}\right)+\frac{\nabla^{2}\sqrt{\rho}}{\sqrt{\rho}}.
\end{equation}
It is well known that the interaction strength in a BEC can be modulated
with an external magnetic field around a Feshbach resonance (see for
example \cite{Schneider2012}). Defining
\begin{equation}
\lambda=\lambda_{0}+\epsilon\lambda_{1}
\end{equation}
and
\begin{equation}
V=V_{0}+\epsilon V_{1}
\end{equation}
for ``unperturbed'' interaction strength $\lambda_{0}$ and external
potential $V_{0}$, we must have
\begin{equation}
V_{1}=-\lambda_{1}\rho.
\end{equation}
From the definitions of $c_{s}$ and $c_{0}$ in Eqs. (\ref{eq:csdef})
and (\ref{eq:c0def}) it is straightforward to show that the interaction
strength perturbation must have the form
\begin{equation}
\lambda_{1}=\frac{\lambda_{0}}{r_{0}}h_{00}
\end{equation}
and thus
\begin{equation}
V_{1}=-\frac{\lambda_{0}\rho}{r_{0}}h_{00},
\end{equation}
where
\begin{equation}
r_{0}=1-\frac{c_{s0}^{2}}{c^{2}}.
\end{equation}
The density may still vary over space, with a shape determined by
the ``unperturbed'' external potential $V_{0}$ as usual. Physically,
this would result in a BEC cloud ``not moving'' in time (no flows,
fixed density distribution), but with a carefully balanced trapping
potential and applied magnetic field changing the speed of sound.
This is somewhat analogous to modulating the refractive index in a
dielectric, a scheme which has also been explored for its applications
in analogue gravity (for example, in \cite{Philbin2008,Leonhardt2002}).
It should be noted that implementing the conditions presented in this
Section does not necessarily result in an exact simulation, as the
effective metrics in Eq. (\ref{eq:metricsimresult}) and Eq. (\ref{eq:directgoal}),
with the above speed of sound perturbation, differ by a conformal
factor, as in the simulation of various black hole geometries in \cite{Fagnocchi2010,Dey2016,Richarte2017}.
Explicitly, in our case,
\begin{equation}
\begin{split}G_{\mu\nu}^{\left(SIM\right)} & =\frac{\rho c}{c_{s}}G_{\mu\nu}^{\left(GW\right)}\\
 & =\frac{\rho c}{c_{s0}}\left(1-\frac{\epsilon}{2}h_{00}+\mathcal{O}\left(\epsilon^{2}\right)\right)G_{\mu\nu}^{\left(GW\right)}.
\end{split}
\end{equation}
The conformal factor will be discussed further in Section \ref{subsec:conformal}.
While a GW is a coordinate independent physical effect, the simulation
presented here reproduces elements of a metric in a particular coordinate
system, and is thus not a coordinate independent solution.

\subsubsection{Non-relativistic limit \label{subsec:nrlimit1}}

In the explicitly non-relativistic limit, the spatial flows are much slower than
the speed of light, so $u^0\rightarrow c$. The interaction
strength must also be weak, so $c_0\ll c$. From the definition of $c_s$
in Eq. (\ref{eq:csdef}), it is clear that we must have $c_s\ll c$. In this
regime, the phononic regime dispersion relation condition in Eq. (\ref{eq:disp})
reduces to
\begin{equation}
\left|k\right|\ll\frac{\sqrt{2}}{\xi}.
\end{equation}
We also assume that the term $\hbar^2\partial_t^2\phi/mc^2$ can
be neglected, i.e. the excitation energy of each boson is much smaller than
its mass energy. The equation governing the evolution of the field
$\phi$ becomes the Gross-Pitaevskii equation,
\begin{equation}
i\hbar\partial_{t}\phi=\left(-\frac{\hbar^{2}}{2m}\nabla^{2}+V^{NR}+\lambda^{NR}\left|\phi\right|^{2}\right)\phi
\end{equation}
where the external potential and interaction strength are related
to those defined in Section \ref{subsec:effmetric} by
\begin{equation}
V^{NR}=\frac{\hbar^{2}}{2m}V,
\end{equation}
\begin{equation}
\lambda^{NR}=\frac{\hbar^{2}}{2m}\lambda.
\end{equation}
In the non-relativistic limit, these are given by
\begin{equation}
V_{1}^{NR}=\epsilon\frac{mc_{s0}^{2}}{2c^{2}}\left(2xy\partial_{t}^{2}h_{\times}+\left(x^{2}-y^{2}\right)\partial_{t}^{2}h_{+}\right),
\end{equation}
\begin{equation}
\lambda_{1}^{NR}=-\epsilon\frac{mc_{s0}^{2}}{2\rho c^{2}}\left(2xy\partial_{t}^{2}h_{\times}+\left(x^{2}-y^{2}\right)\partial_{t}^{2}h_{+}\right).
\end{equation}

\subsection{GW effect simulation \label{subsec:gweffectsim}}

The goal of this section is to simulate the acoustic metric given
in \cite{Sabin2014}, to test the metrological scheme proposed in
\cite{Sabin2014,Sabin2016}. This metric has the form
\begin{equation}
G_{\mu\nu}^{\left(gw\right)}=\frac{\rho_{0}c}{c_{s0}}\left(\eta_{\mu\nu}+h_{\mu\nu}+r_{0}\frac{v_{0\mu}v_{0\nu}}{c^{2}}\right),
\end{equation}
where $\rho_{0}$, $c_{s0}$ and $v_{0}$ are the properties of the
mean field $\phi$ ``unperturbed'' by a simulated GW. In \cite{Sabin2014},
the BEC is considered to be at rest ($v_{0\mu}=-c\delta_{\mu}^{0}$)
in the TT frame, so this is the condition that we will simulate with.
Such a simulation can also be done with a BEC at rest in Fermi normal
coordinates, which is not the initial condition considered in \cite{Sabin2014},
and this solution is presented in the Appendix.

\subsubsection{Acoustic metric with background GW \label{subsec:gwtransform}}

The metric perturbation $h$ in Fermi normal coordinates is given in
Eq. (\ref{eq:gwmetric-2}). Going from the TT frame to Fermi normal
coordinates, the flow velocity transforms as
\begin{equation}
\begin{split}
v_{TT}^{\mu}&=c\delta_{0}^{\mu}\\
\rightarrow v^{\mu}&=c\left(1-\frac{\epsilon}{2}h_{00},-\frac{\epsilon}{2}\partial_{x}H_{00},-\frac{\epsilon}{2}\partial_{y}H_{00},0\right).
\end{split}
\end{equation}
Hence, the acoustic metric has the form
\begin{widetext}
\begin{equation}
G_{\mu\nu}^{\left(gw\right)}=\frac{\rho_{0}c}{c_{s0}}\begin{pmatrix}-c_{s0}^{2}/c^{2}-\epsilon\left(1+r_{0}\right)h_{00} & \frac{r_{0}\epsilon}{2}\partial_{x}H_{00} & \frac{r_{0}\epsilon}{2}\partial_{y}H_{00} & 0\\
\frac{r_{0}\epsilon}{2}\partial_{x}H_{00}\\
\frac{r_{0}\epsilon}{2}\partial_{y}H_{00} &  & \mathbb{I}_{3}\\
0
\end{pmatrix}.\label{eq:gweffmetric}
\end{equation}
\end{widetext}
This is the form of the acoustic metric that we will simulate.

\subsubsection{Acoustic metric with simulated GW}

To simulate the metric in Eq. (\ref{eq:gweffmetric}), we perturb
the bulk properties of the BEC in Fermi normal coordinates (i.e. lab
frame). When constructing the simulation, we consider the background
metric to be the flat Minkowski metric in Eq. (\ref{eq:minkowski}),
with no GW. Let the density, speed of sound and flow velocity be respectively
described as
\begin{align}
\rho&=\rho_{0}+\epsilon\rho_{1},\\
c_{s}^{2}&=c_{s0}^{2}+\epsilon c_{s1}^{2},\label{eq:csdef2}
\end{align}
and
\begin{equation}
v=\left(v_{0},\epsilon v_{1},\epsilon v_{2},0\right),
\end{equation}
with normalisation
\begin{equation}
\left|u\right|=\left|u\right|_{0}+\epsilon\left|u\right|_{1}.
\end{equation}
$\rho_{0}$, $c_{s0}$, $\left|u\right|_{0}$ and $v_{0}$ are bulk
properties of the BEC in the absence of a simulated gravitational
wave, as above. It should be noted that these bulk properties are
not necessarily constant in space or time; they are just the natural
evolution of the BEC with no simulated gravitational wave disturbing
them. Checking the normalisation in Eq. (\ref{eq:vnorm}), we see
that
\begin{equation}
v_{0}=\sqrt{c^{2}-\epsilon^{2}\left(v_{1}^{2}+v_{2}^{2}\right)}=c+\mathcal{O}\left(\epsilon^{2}\right)
\end{equation}
so to first order, $v_{0}=c$. With this flow velocity, the acoustic
metric is
\begin{equation}
G_{\mu\nu}^{\left(sim\right)}=\frac{\rho c}{c_{s}}\begin{pmatrix}-c_{s}^{2}/c^{2} & \epsilon rv_{1}/c & \epsilon rv_{2}/c & 0\\
\epsilon rv_{1}/c\\
\epsilon rv_{2}/c &  & \mathbb{I}_{3}\\
0
\end{pmatrix}\label{eq:simeffmetric}
\end{equation}
to first order in $\epsilon$. Comparison of Eq. (\ref{eq:gweffmetric}) and Eq. (\ref{eq:simeffmetric})
suggests that the velocity perturbation functions should take the
form
\begin{align}
v_{1}&=\frac{1}{2}\partial_{x}H_{00},\label{eq:v1}\\
v_{2}&=\frac{1}{2}\partial_{y}H_{00}.\label{eq:v2}
\end{align}
and the speed of sound perturbation must be
\begin{equation}
c_{s1}^{2}=c^{2}\left(1+r_{0}\right)h_{00}.
\end{equation}
As in Section \ref{subsec:gwmetricsim}, these conditions result in
a conformal simulation, with the conformal factor
\begin{equation}
G_{\mu\nu}^{\left(sim\right)}=\left(1+\epsilon\frac{\rho_{1}}{\rho_{0}}\right)G_{\mu\nu}^{\left(gw\right)}.
\end{equation}
If all components of both effective metrics are to match, there is
no way to avoid this conformal factor. The form of the density perturbation
$\rho_{1}$ required to implements Eqs. (\ref{eq:v1}) and (\ref{eq:v2})
will be calculated in section \ref{subsec:simbulk}. The conformal
factor will be discussed further in section \ref{subsec:conformal}.

\subsubsection{Bulk properties for simulation \label{subsec:simbulk}}

To implement the normalised velocity profile given above, we must
calculate the restrictions placed on the other bulk properties of
the condensate. From Eq. (\ref{eq:irrotational}), we can derive the
velocity normalisation required for irrotational flow. We find that
\begin{equation}
\left|u\right|\left(t,\boldsymbol{x}\right)=\left|u\right|_{0}-\frac{\epsilon}{2c}\partial_{t}\left(\left|u\right|_{0}H_{00}\right),
\end{equation}
where $\boldsymbol{x}$ represents all spatial dimensions ($x$, $y$
and $z$). As in Section \ref{subsec:gwmetricsim}, in the limit of
$\epsilon\rightarrow0$, the BEC is stationary. Hence, the chemical
potential is constant, so $\left|u\right|_{0}$ is constant in both
space and time. Similarly, from these results and the continuity equation,
Eq. (\ref{eq:continuity}), we can derive the form of the density
and its perturbation. We find that the density and its perturbation
have the form
\begin{equation}
\rho_{0}\left(t,\boldsymbol{x}\right)=\frac{\alpha\left(\boldsymbol{x}\right)}{\left|u\right|_{0}}
\end{equation}
and
\begin{equation}
\begin{split}
\rho_{1}&=-\rho_{0}\left(\frac{\left|u\right|_{1}}{\left|u\right|_{0}}\right.\\
&+\left.\frac{c}{2}\left[\left(xh_{+}+yh_{\times}\right)\frac{\partial_{x}\rho_{0}}{\rho_{0}}+\left(xh_{\times}-yh_{+}\right)\frac{\partial_{y}\rho_{0}}{\rho_{0}}\right]\right).
\end{split}
\end{equation}
where $\alpha\left(\boldsymbol{x}\right)$ is some arbitrary function
of integration, encoding the spatial shape of the BEC cloud. It should
be noted that the results of this section are not fundamental restrictions
on the bulk properties on the BEC; rather, they are conditions that
must be imposed in an experiment to facilitate the implementation
of the desired flow velocities and speed of sound. As in Section \ref{subsec:gwmetricsim},
all of these conditions cannot be satisfied with an arbitrary interaction
strength $\lambda$. Taking the same approach as above, we define
a ``perturbed'' interaction strength
\begin{equation}
\lambda=\lambda_{0}+\epsilon\lambda_{1}.
\end{equation}
In general from Eqs. (\ref{eq:csdef}) and (\ref{eq:c0def}), we must
have
\begin{equation}
\lambda_{1}=\lambda_{0}\left(2\frac{\left|u\right|_{1}}{\left|u\right|_{0}}-\frac{\rho_{1}}{\rho_{0}}+\frac{c_{s1}^{2}}{r_{0}c_{s0}^{2}}\right).
\end{equation}
With the results of this section, this expression becomes
\begin{equation}
\begin{split}
\lambda_{1}&=\lambda_{0}\left[\vphantom{\left(\frac{r_{0}}{c_{s0}^{2}}\right)}\right.\left(\frac{c^{2}}{c_{s0}^{2}}\left[1+\frac{1}{r_{0}}\right]-\frac{3}{2}\right)h_{00}\\
 & \left.-\frac{1}{2}\left(\left(xh_{+}+yh_{\times}\right)\frac{\partial_{x}\rho_{0}}{\rho_{0}}+\left(xh_{\times}-yh_{+}\right)\frac{\partial_{y}\rho_{0}}{\rho_{0}}\right)\vphantom{\left(\frac{r_{0}}{c_{s0}^{2}}\right)}\right].
\end{split}
\end{equation}
Using Eq. (\ref{eq:u=00003Dc+stuff}), this corresponds to an external
potential
\begin{equation}
\begin{split}
V=V_{0}+\epsilon\left\{\vphantom{\frac{m^2}{h^2}}\right.-&\lambda_{0}\rho_{1}-\lambda_{1}\rho_{0}+\frac{2m^{2}}{\hbar^{2}}\left|u\right|_{0}\left|u\right|_{1}\\
&\left.+\frac{1}{\sqrt{\rho_{0}}}\left[\frac{\square\sqrt{\rho_{0}}}{\sqrt{\rho_{0}}}+\square\right]\frac{\rho_{1}}{2\sqrt{\rho_{0}}}\right\} .
\end{split}
\end{equation}

\subsubsection{Static bulk solution \label{subsec:staticsolution}}

Consider a BEC trapped in a uniform box potential with infinite potential
walls. In such a case, the density of the BEC is approximately constant
in space everywhere inside the box, apart from a region close to the
boundaries of the trap, where the density goes to zero. The width
of this boundary region is given by the healing length defined above
in Eq. (\ref{eq:healing}). However, as stated in the motivation for
the definition of Eq. (\ref{eq:healing}), we are interested in perturbations
whose wavelength far exceeds the healing length. Hence, for the perturbations
we are considering, in a uniform box potential, we can assume constant
density everywhere. This is also assumed in the detector proposal
\cite{Sabin2014}. As in Section \ref{subsec:gwmetricsim}, it seems
most reasonable to require that $\rho_{0}$ is constant in time, and
thus $\left|u\right|_{0}$ is also. In this case, the perturbed bulk
properties required to simulate a gravitational wave derived above
can be simplified somewhat. Applying these conditions, we find
\begin{align}
\left|u\right|&=\left|u\right|_{0}\left(1-\frac{\epsilon}{2}h_{00}\right),\\
\rho&=\rho_{0}\left(1+\frac{\epsilon}{2}h_{00}\right),
\end{align}
implemented with
\begin{align}
\lambda_{1}&=\lambda_{0}\left(\frac{c^{2}}{c_{s0}^{2}}\left[1+\frac{1}{r_{0}}\right]-\frac{3}{2}\right)h_{00},\\
V_{1}&=-\left[\lambda_{0}\rho_{0}\left(\frac{c^{2}}{c_{s0}^{2}}\left[\frac{3}{2}+\frac{1}{r_{0}}\right]-\frac{3}{2}\right)+\frac{1}{4c^{2}}\partial_{t}^{2}\right]h_{00}.
\end{align}
Using the definitions and conditions presented in Section \ref{subsec:nrlimit1},
the non-relativistic limit of these potential and interaction strength
perturbations are
\begin{align}
\lambda_{1}^{NR}&=-\frac{\lambda_{0}^{NR}}{c_{s0}^{2}}\left(2xy\partial_{t}^{2}h_{\times}+\left(x^{2}-y^{2}\right)\partial_{t}^{2}h_{+}\right),\\
V_{1}^{NR}&=\frac{5\lambda_{0}^{NR}\rho_{0}}{2c_{s0}^{2}}\left(2xy\partial_{t}^{2}h_{\times}+\left(x^{2}-y^{2}\right)\partial_{t}^{2}h_{+}\right).
\end{align}

\subsection{Conformal factor \label{subsec:conformal}}

While the evolution of the phonon field is not generally conformally
invariant in $3+1$ dimensions, there are conformally invariant properties
that may be usefully measured and compared against theoretical predictions.
The Weyl tensor is one of the standard examples of conformally invariant
objects in the framework of general relativity, and for a GW spacetime
has the form
\begin{equation}
C_{\alpha\mu\beta\nu}=-k_{\left[\alpha\right.}\bar{h}_{\left.\mu\right]\left[\nu\right.}k_{\left.\beta\right]},
\end{equation}
where $k_{\mu}$ is the wave-vector of the GW, and $\bar{h}_{\mu\nu}$
is the trace reversed perturbation defined as
\begin{equation}
\bar{h}_{\mu\nu}=h_{\mu\nu}-\frac{1}{2}\eta_{\mu\nu}h_{\hphantom{\sigma}\sigma}^{\sigma}
\end{equation}
for metric perturbation $h_{\mu\nu}$ and Minkowski metric $\eta_{\mu\nu}$
as defined above. This follows simply from the Riemann tensor for
a GW spacetime \cite{MTW}
\begin{equation}
R_{\alpha\mu\beta\nu}=\frac{1}{2}\left(h_{\alpha\nu,\mu\beta}+h_{\mu\beta,\nu\alpha}-h_{\mu\nu,\alpha\beta}-h_{\alpha\beta,\mu\nu}\right).
\end{equation}
In the TT gauge, the elements of the Weyl tensor have simple forms
such as
\begin{equation}
C_{0101}=-k_{z}^{2}h_{+}\,,\,C_{0102}=-k_{z}^{2}h_{\times},
\end{equation}
which can be measured by the detector and compared against experimental
parameters of the simulation.

\section{Examples of GW sources \label{sec:examples}}

\subsection{Non-axisymmetric neutron star}

Rotating neutron stars are one of the strongest predicted sources
of continuous GWs \cite{Aasi2014a}. Any imperfections in the symmetry
of the mass distribution of a neutron star generate gravitational
radiation as the star spins. The simplest case of a non-axisymmetric
neutron star spinning down has strain functions of the form \cite{Palomba2012}
\begin{align}
\epsilon h_{+}\left(t\right)&=h_{0}\left(\frac{1+\cos^{2}\iota}{2}\right)\cos\Phi\left(t\right),\\
\epsilon h_{\times}\left(t\right)&=h_{0}\cos\iota\sin\Phi\left(t\right),
\end{align}
where $\iota$ is the inclination of the neutron star's rotation axis
to the line of sight, the phase evolution is
\begin{equation}
\Phi\left(t\right)=\Phi_{0}+2\pi f\left(t-t_{0}\right)
\end{equation}
for rotation frequency $f/2$ and reference time $t_{0}$, and the
amplitude $h_{0}$ is
\begin{equation}
h_{0}=\frac{4\pi^{2}G}{c^{4}}\frac{I_{zz}\epsilon_{xy}f^{2}}{d}
\end{equation}
with ellipticity
\begin{equation}
\epsilon_{xy}=\frac{I_{xx}-I_{yy}}{I_{zz}}
\end{equation}
where $I_{ii}$ is the moment of inertia of the neutron star about
some $i$ axis, $d$ is the distance to the neutron star and $G$
is Newton's gravitational constant. This coordinate system is defined
such that the axis of rotation is parallel to the $z$ axis. On the
time scale of a detection event, the frequency is constant to very
good approximation, so terms in $\partial_{t}f$ in the phase are
ignored \cite{Palomba2012}. The signal emitted by such a neutron
star can be directly simulated with the interaction and external potential
perturbations
\begin{align}
\lambda_{1}&=\frac{\lambda_{0}f^{2}h_{0}}{2r_{0}c^{2}}\left(\vphantom{\frac{a^2}{2}}2xy\cos\iota\sin\Phi\right.\nonumber\\
&\hphantom{=}+\left.\left(x^{2}-y^{2}\right)\left(\frac{1+\cos^{2}\iota}{2}\right)\cos\Phi\right),\\
V_{1}&=-\lambda_{1}\rho.
\end{align}

\subsection{Compact binary coalescence}

The first direct experimental proof of the existence of GWs was recently
reported by the LIGO collaboration in \cite{Abbott2016}, with the
measurement of the GW signature of the final moments of a compact
binary inspiral involving two black holes. These black holes were
approximately $29$ and $36$ times the mass of the sun respectively,
and $3$ solar masses in energy was radiated in the form of GWs in
the inspiral and collision. The form of the emitted gravitational
radiation during the collision in the ``strong gravity regime''
must be calculated numerically, but the radiation emitted during the
well separated inspiral phase, and the ringdown after coalescence,
has well known solutions.

\subsubsection{Inspiral}

During the inspiral of a compact binary system, while the two compact
objects are still well separated, the gravitational radiation far
from the binary system has the form \cite{MTW}
\begin{align}
\epsilon h_{+}\left(t\right)&=2\left(1+\cos^{2}\iota\right)\frac{\mu}{d}\left[\pi Mf\left(t\right)\right]^{2/3}\cos\left[2\pi F\left(t\right)\right],\\
\epsilon h_{\times}\left(t\right)&=4\cos\iota\frac{\mu}{d}\left[\pi Mf\left(t\right)\right]^{2/3}\sin\left[2\pi F\left(t\right)\right],
\end{align}
where $\iota$ is the inclination axis of inspiral axis to detector,
$M=M_{1}+M_{2}$ and $\mu=M_{1}M_{2}/M^{2}$ for the two masses $M_{1}$
and $M_{2}$, $d$ is the distance from the inspiral barycentre to
the detector,
\begin{equation}
F\left(t\right)=\int^{t}f\left(t'\right)dt',
\end{equation}
and
\begin{equation}
f\left(t\right)=\frac{1}{\pi}\left[\frac{5}{256}\frac{1}{\mu M^{2/3}}\frac{1}{\left(t_{0}-t\right)}\right]^{3/8}
\end{equation}
with some reference time $t_{0}$. This is a sinusoidal signal whose
amplitude and frequency increase as the time $t$ reaches the reference
time $t_{0}$, i.e. the time of collision. This is the characteristic
``chirp'' observed by the LIGO collaboration in \cite{Abbott2016,Abbott2016a,Abbott2017}.
To directly simulate this metric, the corresponding interaction and
external potential perturbations
\begin{widetext}
\begin{align}
\lambda_{1}&=-\frac{\lambda_{0}}{r_{0}c^{2}}\frac{\mu}{d}\left(\pi Mf\right)^{2/3}\left\{4xy\cos\iota\left[\left(\frac{7\pi f}{4\left(t_{0}-t\right)}\right)\cos\left(2\pi F\right)+\left(\frac{5}{16\left(t_{0}-t\right)^{2}}-\left(2\pi f\right)^{2}\right)\sin\left(2\pi F\right)\right]\right.\nonumber\\
& \hphantom{=-\frac{\lambda_{0}}{r_{0}c^{2}}\frac{\mu}{d}}+\left.\left(x^{2}-y^{2}\right)\left(1+\cos^{2}\iota\right)\left[\left(\frac{5}{16\left(t_{0}-t\right)^{2}}-\left(2\pi f\right)^{2}\right)\cos\left(2\pi F\right)-\left(\frac{7\pi f}{4\left(t_{0}-t\right)}\right)\sin\left(2\pi F\right)\right]\right\},\\
V_{1}&=-\lambda_{1}\rho.
\end{align}
\end{widetext}

\subsubsection{Ringdown}

After a binary system with sufficient mass to form a black hole has
collided and coalesced, the resulting black hole rotates due to conservation
of angular momentum. The ringdown of the coalesced object into a stable
rotating black hole can thus be modelled as a perturbed Kerr black
hole. The simplest single-mode ringdown of a Kerr black hole has strain
functions of the form \cite{Aasi2014}
\begin{equation}
\epsilon h_{+}\left(t\right)=\frac{\mathcal{A}}{d}\left(1+\cos^{2}\iota\right)e^{\left(\Phi_{0}-\Phi\left(t\right)\right)/2Q}\cos\Phi\left(t\right),
\end{equation}
\begin{equation}
\epsilon h_{\times}\left(t\right)=\frac{\mathcal{A}}{d}\left(2\cos\iota\right)e^{\left(\Phi_{0}-\Phi\left(t\right)\right)/2Q}\sin\Phi\left(t\right),
\end{equation}
where
\begin{equation}
\Phi\left(t\right)=\Phi_{0}+2\pi f\left(t-t_{0}\right)
\end{equation}
as above, $\iota$ is still the inclination angle of rotation axis
to the detector, $d$ is the distance from the source to the detector,
$Q$ is the ``quality factor'' fitted numerically with
\begin{equation}
Q=0.7000+1.4187\left(1-\hat{a}\right)^{-0.4990}
\end{equation}
for spin parameter $\hat{a}=cS/GM^{2}$, with spin angular momentum
$S$. The GW amplitude $\mathcal{A}$ is given by
\begin{equation}
\mathcal{A}=\frac{GM}{c^{2}}\sqrt{\frac{5\varepsilon}{2}}Q^{-1/2}F\left(Q\right)^{-1/2}g\left(\hat{a}\right)^{-1/2}
\end{equation}
where $F\left(Q\right)=1+1/4Q^{2}$, $g\left(\hat{a}\right)=1.5251-1.1568\left(1-\hat{a}\right)^{0.1292}$
and $\varepsilon$ is the fraction of the black hole mass radiated
away. Functionally, this is a decaying sinusoid of constant frequency.
The corresponding interaction and external potential perturbations
for simulation are
\begin{widetext}
\begin{align}
\lambda_{1}&=-\frac{\lambda_{0}}{r_{0}c^{2}}\left\{ 2xy\cos\iota\left(\left[\frac{1}{4Q^{2}}-1\right]\sin\Phi-\frac{1}{Q}\cos\Phi\right)\right.\nonumber\\
& \hphantom{=-\frac{\lambda_{0}}{r_{0}c^{2}}}\left.+\frac{1}{2}\left(x^{2}-y^{2}\right)\left(1+\cos^{2}\iota\right)\left[\frac{1}{Q}\sin\Phi+\left(\frac{1}{4Q^{2}}-1\right)\cos\Phi\right]\right\} \frac{\mathcal{A}}{d}\left(2\pi f\right)^{2}e^{\left(\Phi_{0}-\Phi\right)/2Q},\\
V_{1}&=-\lambda_{1}\rho.
\end{align}
\end{widetext}

\section{Reduction to 1+1 \label{sec:1+1}}

In this Section, we restrict ourselves to an effective $1$-dimensional
field to compare to earlier work in \cite{Bravo2015}. In an effective
$1+1$ dimensional spacetime, the GW metric reduces to
\begin{equation}
g_{\mu\nu}=\begin{pmatrix}-1-\epsilon x^{2}\partial_{t}^{2}h_{+}/2c^{2} & 0\\
0 & 1
\end{pmatrix}
\end{equation}
To simulate this, the speed of sound is chosen as
\begin{equation}
c_{s}^{2}=c_{s0}^{2}\left(1-\epsilon\frac{x^{2}}{2}\partial_{t}^{2}h_{+}\right)
\end{equation}
and a time scaling of
\begin{equation}
c_{s0}\tau=\left(\frac{c_{s0}^{2}}{c^{2}}\right)ct
\end{equation}
results in a simulation
\begin{equation}
G_{\mu'\nu'}^{\left(SIM\right)}=\frac{\rho c}{c_{s}}\begin{pmatrix}-1-\epsilon x^{2}\partial_{t}^{2}h_{+}/2c^{2} & 0\\
0 & 1
\end{pmatrix}.
\end{equation}
In $1+1$ dimensions, the equations of motion are conformally independent,
so this is an exact simulation. Following the same procedure as in
Section \ref{subsec:gwmetricsim}, we require that the flow velocity
normalisation is completely constant and conclude that the density
is constant in time. The interaction strength and external potential
perturbations required to implement this are then
\begin{equation}
\lambda_{1}=-\frac{\lambda_{0}}{r_{0}c^{2}}\left(\frac{x^{2}}{2}\partial_{t}^{2}h_{+}\right),
\end{equation}
\begin{equation}
V_{1}=\frac{\lambda_{0}\rho}{r_{0}c^{2}}\left(\frac{x^{2}}{2}\partial_{t}^{2}h_{+}\right).
\end{equation}
We must stress that this is an effective $1+1$ dimensional theory,
and care must be taken when dealing with the actual field dynamics.
Although this seems to work at the level of the metric, a naive suppression
of the remaining spatial dimensions cannot be done due to the fact
that the conformal factor is dimensionally dependent, and diverges
when the number of spatial dimensions is exactly $1$ \cite{Barcelo2005}.
Nevertheless, as long as the system is sufficiently constrained in
the extra dimensions, e.g. in a highly elongated trap, a well-behaved
effective $1+1$ dimensional system can always be constructed.

\section{Conclusion \label{sec:conclusion}}

We have shown how to simulate a GW spacetime in $3+1$ dimensions
for quantum excitations of a BEC, up to a conformal factor, as well
as simulating the acoustic metric used in \cite{Sabin2014} to propose
a GW detector. By making use of the ``gauge freedom'' of the GW
metric corresponding to a linearised coordinate transformation, we
chose a frame in which the metric perturbation could be simulated
by perturbing the speed of sound in the BEC. We then examined the
restrictions this places on other bulk properties through the continuity
equation and experimental limitations, and calculated the external
and interaction potential perturbations needed to implement such a
simulation in the lab. Although the simulated metric is related to
the target metric by a non-constant conformal factor, we show that
there are still useful properties that can be measured and tested
in an experiment. We also give explicit expressions for the simulation
of GWs from various sources. This work generalises the results of
\cite{Bravo2015} and presents a complementary approach to simulation
in effectively $1+1$ dimensional BECs.

The results presented here can also be derived in the context of an
explicitly non-relativistic treatment of a BEC, such as that derived
in \cite{Barcelo2005}. In a non-relativistic BEC, phonons on the
BEC still propagate on a Lorentzian effective spacetime described
by an acoustic metric, but this metric is necessarily spatially conformally
flat. We consider a BEC in a covariant formalism in this paper to
match the approach of \cite{Sabin2014,Sabin2016,Howl2016a} for the
simulation in Section \ref{subsec:gweffectsim}, and to express the
interaction of a BEC with GWs in a natural way. As explained in Section
\ref{subsec:effmetric}, the simulation of GWs presented in this paper
do not rely on the relativistic nature of the BEC or any relativistic effects,
nor do the perturbations to the external and interaction potentials
disappear in the non-relativistic limit.

We have studied GWs in the context of perturbations around a flat
spacetime metric and assuming GWs to be far outside the source. Other
interesting simulations could involve GWs propagating on curved backgrounds,
such as black holes \cite{Cropp2016,Dey2016} or during inflation
\cite{Fedichev2003}, or in strong-field regimes. Furthermore, since
phonons are quantum quasi-particles, this opens up the possibility
of studying predicted effects of quantum field theory in curved spacetime,
such as how a GW may affect the entanglement of quantum systems, a
phenomena that is utilised in the BEC GW detector proposed in \cite{Sabin2014}.
This, therefore, also presents a potential, and fully configurable,
testing environment for this GW detector metrological scheme. To obtain
a full simulation of the GW detector, we need a better understanding
of the effect of the GW on the bulk of the BEC. As mentioned in the conclusion
of \cite{Bravo2015}, an experimental simulation of the effect of a large
amplitude GW and subsequent detection of phonons would also be a
proof-of-concept demonstration of the generation of phonons by GWs
as predicted in \cite{Sabin2014}.

\begin{acknowledgments}
The authors thank Joel Lindkvist and David Edward Bruschi for helpful
discussion and comments. D.H. acknowledges funding from CoQuS. T.B.
acknowledges funding from CONACYT under project code 261699/359033.
D.R. acknowledges funding from the Humboldt Foundation. R.H. and I.F.
would like to acknowledge that this project was made possible through
the support of the grant \textquoteleft Leaps in cosmology: gravitational
wave detection with quantum systems\textquoteright{} (No. 58745) from
the John Templeton Foundation. The opinions expressed in this publication
are those of the authors and do not necessarily reflect the views
of the John Templeton Foundation.
\end{acknowledgments}

\appendix

\section{Simulation in alternative coordinates}

As explained in Section \ref{subsec:gweffectsim}, it is possible
to simulate the effect of a GW starting with no flows in Fermi normal
coordinates. This solution is presented here. As above, let the density,
speed of sound and flow velocity normalisation be respectively described
as
\begin{equation}
\rho=\rho_{0}+\epsilon\rho_{1},
\end{equation}
\begin{equation}
c_{s}^{2}=c_{s0}^{2}+\epsilon c_{s1}^{2},
\end{equation}
and
\begin{equation}
\left|u\right|=\left|u\right|_{0}+\epsilon\left|u\right|_{1},
\end{equation}
where $\epsilon$ is the small parameter defined above. $\rho_{0}$,
$c_{s0}$, and $\left|u\right|_{0}$ are bulk properties of the BEC
in the absence of a simulated GW. 

\subsection{Effective metric with background gravitational wave}

In Fermi normal coordinates, we consider the flows in the BEC to be
$v_{\mu}=-v_{0}\delta_{\mu}^{0}$. With the normalisation equation
(Eq. (\ref{eq:vnorm})), we can determine the function $v_{0}$ as
\[
g^{\mu\nu}v_{\mu}v_{\nu}=-c^{2}
\]
\begin{equation}
\begin{split}g^{\mu\nu}v_{\mu}v_{\nu} & =\left(\eta^{\mu\nu}-\epsilon h^{\mu\nu}\right)v_{0}\delta_{\mu}^{0}v_{0}\delta_{\nu}^{0}\\
 & =\left(-1+\epsilon h_{00}\right)v_{0}^{2}\\
 & =-c^{2}
\end{split}
\end{equation}
\begin{equation}
\implies v_{0}^{2}=c^{2}\left(1+\epsilon h_{00}\right).
\end{equation}
Then, with a background GW, the acoustic metric has the form
\begin{equation}
G_{\mu\nu}^{\left(gw\right)}=\frac{\rho_{0}c}{c_{s0}}\begin{pmatrix}-c_{s0}^{2}/c^{2}\left(1+\epsilon h_{00}\right) & \boldsymbol{0}^{\text{T}}\\
\boldsymbol{0} & \mathbb{I}_{3}
\end{pmatrix}+\mathcal{O}\left(\epsilon^{2}\right).\label{eq:gweffappendix}
\end{equation}

\subsection{Effective metric with simulated gravitational wave}

Consider the background metric $g_{\mu\nu}$ in Fermi normal coordinates
to be the flat Minkowski metric $\eta_{\mu\nu}$ defined as in Eq.
(\ref{eq:minkowski}), with no background GW ($h_{\mu\nu}=0$). In
these coordinates, consider the case where there are no flows on the
BEC, so the flow velocity is
\begin{equation}
\left.v\right|_{h\rightarrow0}=\left(c,0,0,0\right).
\end{equation}
To simulate the effect of a GW, we perturb the bulk properties of
the BEC. The acoustic metric is
\begin{equation}
G_{\mu\nu}^{\left(sim\right)}=\frac{\rho c}{c_{s}}\begin{pmatrix}-c_{s}^{2}/c^{2} & \boldsymbol{0}^{\text{T}}\\
\boldsymbol{0} & \mathbb{I}_{3}
\end{pmatrix}+\mathcal{O}\left(\epsilon^{2}\right).\label{eq:simeffappendix}
\end{equation}

\subsection{Simulation}

Comparison of Eq. (\ref{eq:gweffappendix}) and Eq. (\ref{eq:simeffappendix})
suggests that, to simulate a background GW in these coordinates, the
speed of sound should be modulated as
\begin{equation}
c_{s}^{2}=c_{s0}^{2}\left(1+\epsilon h_{00}\right).
\end{equation}
As with the simulation presented in the main body of the paper, this
simulation differs from an exact simulation by a conformal factor;
\begin{equation}
G_{\mu\nu}^{\left(sim\right)}=\left(1+\epsilon\left[\frac{\rho_{1}}{\rho_{0}}-\frac{h_{00}}{2}\right]\right)G_{\mu\nu}^{\left(gw\right)}.
\end{equation}

\subsubsection{Bulk properties for simulation}

To implement the normalised velocity profile given above, we must
calculate the restrictions placed on the other bulk properties of
the condensate. From Eq. (\ref{eq:irrotational}),
\begin{equation}
\left|u\right|=\left|u\right|_{0}\left(t\right),
\end{equation}
and from Eq. (\ref{eq:continuity}),
\begin{equation}
\rho_{0}\left(t,\boldsymbol{x}\right)=\frac{\alpha\left(\boldsymbol{x}\right)}{\left|u\right|_{0}\left(t\right)}.
\end{equation}
As in Section \ref{subsec:gwmetricsim}, it seems most reasonable
to require that $\partial_{t}\rho_{0}=0$ and so $\left|u\right|_{0}$
is completely constant. Defining a ``perturbed'' interaction strength
as
\begin{equation}
\lambda=\lambda_{0}+\epsilon\lambda_{1},
\end{equation}
the results of this Section and Eqs. (\ref{eq:csdef}) and (\ref{eq:c0def})
are simultaneously satisfied if
\begin{equation}
\lambda_{1}=\frac{\lambda_{0}}{r_{0}}h_{00}.
\end{equation}
This can be implemented together with the external potential
\begin{equation}
V_{1}=-\frac{\lambda_{0}\rho}{r_{0}}h_{00}.
\end{equation}

\section{Acoustic metric with general background metric}
\label{sec:metricderivation}

In \cite{Fagnocchi2010,Visser2010}, the acoustic metric is derived for a
flat Minkowski background metric. Since we require the same for a general
background metric, in this appendix we extend the acoustic metric
to the case where the background metric is not necessarily flat.

\subsection{Equations of motion and basic approximations}

\subsubsection{Lagrangian}

The Lagrangian density for an interacting massive complex
scalar field $\hat{\Phi}$ on a (in general curved) background with
metric $g_{\mu\nu}$ may be written as
\begin{equation}
\mathcal{L}=-\sqrt{-g}\left\{ g^{\mu\nu}\partial_{\mu}\hat{\Phi}^{\dagger}\partial_{\nu}\hat{\Phi}+\left(\frac{m^{2}c^{2}}{\hbar^{2}}+V\right)\hat{\Phi}^{\dagger}\hat{\Phi}+U\left(\hat{\Phi}^{\dagger}\hat{\Phi},\lambda_{i}\right)\right\} \label{eq:lag}
\end{equation}
where $m$ is the mass, the external potential $V$ is generally a
function of space and time, and the interaction potential $U$ depends
on coupling constants $\lambda_{i}$ which are also in principle functions
of space and time. The background metric $g_{\mu\nu}$ cannot be
completely general; we restrict ourselves to spacetimes with sufficiently
weak curvature such that Bose-Einstein condensation can still be well defined.
Further restrictions on the metric will be given in Section \ref{restrictionsetc}.
The interaction potential $U$ can be expanded as
\begin{equation}
U\left(\hat{\Phi}^{\dagger}\hat{\Phi},\lambda_{i}\right)=\frac{1}{2!}\lambda_{2}\hat{\Phi}^{\dagger}\hat{\Phi}^{\dagger}\hat{\Phi}\hat{\Phi}+\frac{1}{3!}\lambda_{3}\hat{\Phi}^{\dagger}\hat{\Phi}^{\dagger}\hat{\Phi}^{\dagger}\hat{\Phi}\hat{\Phi}\hat{\Phi}+\cdots.
\end{equation}
We will consider only the first term of $U$ corresponding to two-particle
interactions, and ignore further terms corresponding to three or more
particle interactions. For notational convenience, we will drop the
label on $\lambda_{2}$ so
\begin{equation}
U\left(\hat{\Phi}^{\dagger}\hat{\Phi},\lambda_{i}\right)\approx\frac{1}{2}\lambda\hat{\Phi}^{\dagger}\hat{\Phi}^{\dagger}\hat{\Phi}\hat{\Phi}.
\end{equation}
The Euler-Lagrange equation for $\hat{\Phi}^{\dagger}$ is
\begin{equation}
\left[\square_{g}-\left(\frac{m^{2}c^{2}}{\hbar^{2}}+V\right)-\lambda\hat{\Phi}^{\dagger}\hat{\Phi}\right]\hat{\Phi}=0\label{eq:nlkg}
\end{equation}
where
\begin{equation}
\square_{g}\hat{\Phi}=\frac{1}{\sqrt{-g}}\partial_{\mu}\left(\sqrt{-g}g^{\mu\nu}\partial_{\nu}\hat{\Phi}\right)
\end{equation}
and $g$ is the determinant of $g_{\mu\nu}$.

\subsubsection{Approximations}

We now let this field $\hat{\Phi}$ represent a Bose-Einstein condensate
and make the Bogoliubov approximation to separate the ``condensed
fraction'' of the field $\phi$ from a small ``uncondensed fraction''
$\hat{\psi}$. This is done multiplicatively as
\begin{equation}
\hat{\Phi}=\phi\left(1+\hat{\psi}\right)\label{eq:bogo}
\end{equation}
to simplify the equation for $\hat{\psi}$ later. As part of the Bogoliubov
approximation, we say that
\begin{equation}
\left\langle \hat{\Phi}\right\rangle =\phi\implies\left\langle \hat{\psi}\right\rangle =0
\end{equation}
where $\left\langle \cdot\right\rangle $ is a non-equilibrium average.
Taking the average of Eq. (\ref{eq:nlkg}),
\begin{equation}
\left[\square_{g}-\left(\frac{m^{2}c^{2}}{\hbar^{2}}+V\right)-\lambda\left|\phi\right|^{2}\right]\phi-\lambda\left|\phi\right|^{2}\phi\left[\left\langle \hat{\psi}\hat{\psi}\right\rangle +2\left\langle \hat{\psi}^{\dagger}\hat{\psi}\right\rangle +\left\langle \hat{\psi}^{\dagger}\hat{\psi}\hat{\psi}\right\rangle \right]=0.
\end{equation}
We now take the Popov approximation
\begin{equation}
\left\langle \hat{\psi}\hat{\psi}\right\rangle =0=\left\langle \hat{\psi}\hat{\psi}\hat{\psi}\right\rangle 
\end{equation}
and require that the density of excited atoms is much smaller than
the density of mean-field atoms, i.e.
\begin{equation}
\left\langle \hat{\psi}^{\dagger}\hat{\psi}\right\rangle \ll1.
\end{equation}
This results in a non-linear Klein-Gordon-like equation for the mean
field $\phi$:
\begin{equation}
\left[\square_{g}-\left(\frac{m^{2}c^{2}}{\hbar^{2}}+V\right)-\lambda\left|\phi\right|^{2}\right]\phi=0.\label{eq:meanfield}
\end{equation}
This is a curved space-time generalisation of the Gross-Pitaevskii equation.
In flat space-time (where the metric is the Minkowski metric $\eta_{\mu\nu}$
defined in Eq. (\ref{eq:minkowski})) and in the non-relativistic limit, we can replace
$\phi$ with a lower energy field
\begin{equation}
\phi=\varphi e^{imc^{2}t/\hbar}
\end{equation}
and take the limit of $c\rightarrow\infty$. Assuming that the energy
of excitations in $\varphi$ is sufficiently low such that we can ignore
terms of order $\partial_{t}^{2}\varphi$, the remaining terms of
Eq. (\ref{eq:meanfield}) have the form
\begin{equation}
i\hbar\partial_{t}\varphi=\left[-\frac{\hbar^{2}}{2m}\nabla^{2}+V_{NR}+g_{NR}\left|\varphi\right|^{2}\right]\varphi
\end{equation}
where
\begin{equation}
g_{NR}=\frac{\hbar^{2}}{2m}\lambda\,,\,V_{NR}=\frac{\hbar^{2}}{2m}V,
\end{equation}
which is the usual time dependent Gross-Pitaevskii equation.

\subsubsection{Continuity and velocity normalisation equations}

If the mean field $\phi$ is written in the Madelung representation
$\phi=\sqrt{\rho}e^{i\theta}$ and defining a flow velocity
\begin{equation}
u^{\mu}=\frac{\hbar}{m}g^{\mu\nu}\partial_{\nu}\theta,
\end{equation}
then separating the real and imaginary components of Eq. (\ref{eq:meanfield})
results in two equations:
\begin{equation}
\nabla_{\mu}\left(\rho u^{\mu}\right)=0,\label{eq:continuity}
\end{equation}
\begin{equation}
-g_{\mu\nu}u^{\mu}u^{\nu}=c^{2}+\frac{\hbar^{2}}{m^{2}}\left\{ V+\lambda\rho-\frac{\square_{g}\sqrt{\rho}}{\sqrt{\rho}}\right\} .\label{eq:velnorm}
\end{equation}
Eq. (\ref{eq:continuity}) is a continuity equation, and can also
be derived from the global phase $U\left(1\right)$ symmetry of the
Lagrangian density in Eq. (\ref{eq:lag}). Eq. (\ref{eq:velnorm})
allows us to directly relate the external and interaction potentials
with the mean field properties of the BEC without necessarily solving
the full dynamics with Eq. (\ref{eq:meanfield}).

\subsection{Phonon equations}

\subsubsection{Equations for $\hat{\psi}$}

Combining Eqs. (\ref{eq:nlkg}), (\ref{eq:bogo}) and (\ref{eq:meanfield}),
we find
\begin{equation}
\left[i\hbar u^{\mu}\partial_{\mu}-\hat{T}_{\rho}-mc_{0}^{2}\right]\hat{\psi}=mc_{0}^{2}\hat{\psi}^{\dagger}\label{eq:phonons1}
\end{equation}
where
\begin{equation}
c_{0}^{2}=\frac{\hbar^{2}}{2m^{2}}\lambda\rho,
\end{equation}
and
\begin{equation}
\hat{T}_{\rho}\hat{\psi}=-\frac{\hbar^{2}}{2m\rho\sqrt{-g}}\partial_{\mu}\left(\rho\sqrt{-g}g^{\mu\nu}\partial_{\nu}\hat{\psi}\right)
\end{equation}
is a generalised kinetic operator, which reduces to the standard kinetic
energy operator $T=-\left(\hbar^{2}/2m\right)\nabla^{2}$ for constant
$\rho$ in the non-relativistic flat space-time limit. Note that we
require the solution to Eq. (\ref{eq:meanfield}) to solve Eq. (\ref{eq:phonons1})
but not vice-versa, as we are neglecting the back-reaction of $\hat{\psi}$
on $\phi$. Taking the equivalent equation to Eq. (\ref{eq:phonons1})
for $\hat{\psi}^{\dagger}$ and combining these to eliminate $\hat{\psi}^{\dagger}$,
we find
\begin{equation}
\left(\left[i\hbar u^{\mu}\partial_{\mu}+\hat{T}_{\rho}\right]\frac{1}{c_{0}^{2}}\left[-i\hbar u^{\mu}\partial_{\mu}+\hat{T}_{\rho}\right]+2m\hat{T}_{\rho}\right)\hat{\psi}=0.\label{eq:phonons2}
\end{equation}
It is important to note that although Eq. (\ref{eq:phonons1}) implies
Eq. (\ref{eq:phonons2}), the converse is not true.

\subsubsection{Relative term strength}
\label{restrictionsetc}

The phonon equation Eq. (\ref{eq:phonons2}) can be expanded into
four terms as
\begin{equation}
\hat{T}_{1}+\hat{T}_{2}+\hat{T}_{3}+\hat{T}_{4}=0
\end{equation}
where
\begin{equation}
\hat{T}_{1}=i\hbar u^{\mu}\partial_{\mu}\frac{1}{c_{0}^{2}}\left[-i\hbar u^{\nu}\partial_{\nu}\right]\hat{\psi},
\end{equation}
\begin{equation}
\hat{T}_{2}=i\hbar\left[u^{\mu}\partial_{\mu}\frac{1}{c_{0}^{2}}\hat{T}_{\rho}-\hat{T}_{\rho}\frac{1}{c_{0}^{2}}u^{\mu}\partial_{\mu}\right]\hat{\psi},
\end{equation}
\begin{equation}
\hat{T}_{3}=\hat{T}_{\rho}\frac{1}{c_{0}^{2}}\hat{T}_{\rho}\hat{\psi}
\end{equation}
and
\begin{equation}
\hat{T}_{4}=2m\hat{T}_{\rho}\hat{\psi}.
\end{equation}
We make an eikonal approximation, where
\begin{equation}
\left|\frac{\partial_{t}\rho}{\rho}\right|\ll\omega,\,\left|\frac{\partial_{t}c_{0}}{c_{0}}\right|\ll\omega,\,\left|\frac{\partial_{t}u^{\mu}}{u^{\mu}}\right|\ll\omega
\end{equation}
and the corresponding relations for variations in space as in the
flat space case, but also
\begin{equation}
\left|\frac{\partial_{t}g_{\mu\nu}}{g_{\mu\nu}}\right|\ll\omega,\,\left|\frac{\partial_{t}g}{2g}\right|\ll\omega
\label{gllw}
\end{equation}
with the corresponding relations for variations in space. Note that
Eq. (\ref{gllw}) restricts the curvature of the metric with respect to the phonon
mode frequencies. For linearised gravity and realistic phonon frequencies, this
will always hold. Additionally, following \cite{Fagnocchi2010,Bruschi2014}
we consider small momenta in the phononic regime, such that the dispersion
relation is linear and terms quartic in $k$ can be neglected. With these
approximations, $\hat{T}_{2}$ and $\hat{T}_{3}$ are negligible in comparison to $\hat{T}_{1}$
and $\hat{T}_{4}$, so we are left with
\begin{equation}
\left[u^{\mu}\partial_{\mu}\frac{1}{c_{0}^{2}}u^{\nu}\partial_{\nu}+\frac{2m}{\hbar^{2}}\hat{T}_{\rho}\right]\hat{\psi}=0.\label{eq:phonons3}
\end{equation}
Expanding Eq. (\ref{eq:phonons3}), we find an equation of the form
\begin{equation}
\partial_{\mu}\left(f^{\mu\nu}\partial_{\nu}\hat{\psi}\right)=0\label{eq:kg}
\end{equation}
where
\begin{equation}
f^{\mu\nu}=\rho\sqrt{-g}\left[g^{\mu\nu}-\frac{u^{\mu}u^{\nu}}{c_{0}^{2}}\right].
\end{equation}

\subsubsection{Acoustic metric}

Eq. (\ref{eq:kg}) has a form similar to a Klein-Gordon equation for
a massless non-interacting scalar field $\hat{\psi}$ in a space-time
with an effective metric given by
\begin{equation}
f^{\mu\nu}=\sqrt{-G}G^{\mu\nu}.
\end{equation}
Taking the determinant of this equation, we have
\begin{equation}
G=-\left(-f\right)^{\frac{2}{n-2}}
\end{equation}
where $n$ is the total number of dimensions, noting that $f$ is
the determinant of $f^{\mu\nu}$, but $G$ is the determinant of $G_{\mu\nu}$.
Taking the determinant of $f^{\mu\nu}$, we find that
\begin{equation}
\sqrt{-G}=\sqrt{-g}\rho^{\frac{n}{n-2}}\left(\frac{c}{c_{s}}\right)^{\frac{2}{n-2}}
\end{equation}
where the scalar speed of sound $c_{s}$ is defined as
\begin{equation}
c_{s}^{2}=\frac{c^{2}c_{0}^{2}}{\left|u\right|^{2}+c_{0}^{2}}.
\end{equation}
Hence, defining normalised flow velocity as
\begin{equation}
v^{\mu}=\frac{c}{\left|u\right|}u^{\mu},
\end{equation}
the inverse general acoustic metric is
\begin{equation}
G^{\mu\nu}=\left(\frac{\rho c}{c_{s}}\right)^{-\frac{2}{n-2}}\left[g^{\mu\nu}+\left(1-\frac{c^{2}}{c_{s}^{2}}\right)\frac{v^{\mu}v^{\nu}}{c^{2}}\right]
\end{equation}
which can be inverted to define the general acoustic metric
\begin{equation}
G_{\mu\nu}=\left(\frac{\rho c}{c_{s}}\right)^{\frac{2}{n-2}}\left[g_{\mu\nu}+\left(1-\frac{c_{s}^{2}}{c^{2}}\right)\frac{v_{\mu}v_{\nu}}{c^{2}}\right].
\end{equation}


\begin{thebibliography}{43}%
\makeatletter
\providecommand \@ifxundefined [1]{%
 \@ifx{#1\undefined}
}%
\providecommand \@ifnum [1]{%
 \ifnum #1\expandafter \@firstoftwo
 \else \expandafter \@secondoftwo
 \fi
}%
\providecommand \@ifx [1]{%
 \ifx #1\expandafter \@firstoftwo
 \else \expandafter \@secondoftwo
 \fi
}%
\providecommand \natexlab [1]{#1}%
\providecommand \enquote  [1]{``#1''}%
\providecommand \bibnamefont  [1]{#1}%
\providecommand \bibfnamefont [1]{#1}%
\providecommand \citenamefont [1]{#1}%
\providecommand \href@noop [0]{\@secondoftwo}%
\providecommand \href [0]{\begingroup \@sanitize@url \@href}%
\providecommand \@href[1]{\@@startlink{#1}\@@href}%
\providecommand \@@href[1]{\endgroup#1\@@endlink}%
\providecommand \@sanitize@url [0]{\catcode `\\12\catcode `\$12\catcode
  `\&12\catcode `\#12\catcode `\^12\catcode `\_12\catcode `\%12\relax}%
\providecommand \@@startlink[1]{}%
\providecommand \@@endlink[0]{}%
\providecommand \url  [0]{\begingroup\@sanitize@url \@url }%
\providecommand \@url [1]{\endgroup\@href {#1}{\urlprefix }}%
\providecommand \urlprefix  [0]{URL }%
\providecommand \Eprint [0]{\href }%
\providecommand \doibase [0]{http://dx.doi.org/}%
\providecommand \selectlanguage [0]{\@gobble}%
\providecommand \bibinfo  [0]{\@secondoftwo}%
\providecommand \bibfield  [0]{\@secondoftwo}%
\providecommand \translation [1]{[#1]}%
\providecommand \BibitemOpen [0]{}%
\providecommand \bibitemStop [0]{}%
\providecommand \bibitemNoStop [0]{.\EOS\space}%
\providecommand \EOS [0]{\spacefactor3000\relax}%
\providecommand \BibitemShut  [1]{\csname bibitem#1\endcsname}%
\let\auto@bib@innerbib\@empty
\bibitem [{\citenamefont {Unruh}(1981)}]{Unruh1981}%
  \BibitemOpen
  \bibfield  {author} {\bibinfo {author} {\bibfnamefont {W.~G.}\ \bibnamefont
  {Unruh}},\ }\href {\doibase 10.1103/PhysRevLett.46.1351} {\bibfield
  {journal} {\bibinfo  {journal} {Physical Review Letters}\ }\textbf {\bibinfo
  {volume} {46}},\ \bibinfo {pages} {1351} (\bibinfo {year}
  {1981})}\BibitemShut {NoStop}%
\bibitem [{\citenamefont {Lahav}\ \emph {et~al.}(2010)\citenamefont {Lahav},
  \citenamefont {Itah}, \citenamefont {Blumkin}, \citenamefont {Gordon},
  \citenamefont {Rinott}, \citenamefont {Zayats},\ and\ \citenamefont
  {Steinhauer}}]{Lahav2010}%
  \BibitemOpen
  \bibfield  {author} {\bibinfo {author} {\bibfnamefont {O.}~\bibnamefont
  {Lahav}}, \bibinfo {author} {\bibfnamefont {A.}~\bibnamefont {Itah}},
  \bibinfo {author} {\bibfnamefont {A.}~\bibnamefont {Blumkin}}, \bibinfo
  {author} {\bibfnamefont {C.}~\bibnamefont {Gordon}}, \bibinfo {author}
  {\bibfnamefont {S.}~\bibnamefont {Rinott}}, \bibinfo {author} {\bibfnamefont
  {A.}~\bibnamefont {Zayats}}, \ and\ \bibinfo {author} {\bibfnamefont
  {J.}~\bibnamefont {Steinhauer}},\ }\href {\doibase
  10.1103/PhysRevLett.105.240401} {\bibfield  {journal} {\bibinfo  {journal}
  {Physical Review Letters}\ }\textbf {\bibinfo {volume} {105}},\ \bibinfo
  {pages} {240401} (\bibinfo {year} {2010})}\BibitemShut {NoStop}%
\bibitem [{\citenamefont {Hung}\ \emph {et~al.}(2013)\citenamefont {Hung},
  \citenamefont {Gurarie},\ and\ \citenamefont {Chin}}]{Hung2013}%
  \BibitemOpen
  \bibfield  {author} {\bibinfo {author} {\bibfnamefont {C.-L.}\ \bibnamefont
  {Hung}}, \bibinfo {author} {\bibfnamefont {V.}~\bibnamefont {Gurarie}}, \
  and\ \bibinfo {author} {\bibfnamefont {C.}~\bibnamefont {Chin}},\ }\href
  {\doibase 10.1126/science.1237557} {\bibfield  {journal} {\bibinfo  {journal}
  {Science}\ }\textbf {\bibinfo {volume} {341}},\ \bibinfo {pages} {1213}
  (\bibinfo {year} {2013})}\BibitemShut {NoStop}%
\bibitem [{\citenamefont {Steinhauer}(2014)}]{Steinhauer2014}%
  \BibitemOpen
  \bibfield  {author} {\bibinfo {author} {\bibfnamefont {J.}~\bibnamefont
  {Steinhauer}},\ }\href {http://dx.doi.org/10.1038/nphys3104} {\bibfield
  {journal} {\bibinfo  {journal} {Nature Physics}\ }\textbf {\bibinfo {volume}
  {10}},\ \bibinfo {pages} {864} (\bibinfo {year} {2014})}\BibitemShut
  {NoStop}%
\bibitem [{\citenamefont {Weinfurtner}\ \emph {et~al.}(2011)\citenamefont
  {Weinfurtner}, \citenamefont {Tedford}, \citenamefont {Penrice},
  \citenamefont {Unruh},\ and\ \citenamefont {Lawrence}}]{Weinfurtner2011}%
  \BibitemOpen
  \bibfield  {author} {\bibinfo {author} {\bibfnamefont {S.}~\bibnamefont
  {Weinfurtner}}, \bibinfo {author} {\bibfnamefont {E.~W.}\ \bibnamefont
  {Tedford}}, \bibinfo {author} {\bibfnamefont {M.~C.~J.}\ \bibnamefont
  {Penrice}}, \bibinfo {author} {\bibfnamefont {W.~G.}\ \bibnamefont {Unruh}},
  \ and\ \bibinfo {author} {\bibfnamefont {G.~A.}\ \bibnamefont {Lawrence}},\
  }\href {\doibase 10.1103/PhysRevLett.106.021302} {\bibfield  {journal}
  {\bibinfo  {journal} {Physical Review Letters}\ }\textbf {\bibinfo {volume}
  {106}},\ \bibinfo {pages} {021302} (\bibinfo {year} {2011})}\BibitemShut
  {NoStop}%
\bibitem [{\citenamefont {Euv\'e}\ \emph {et~al.}(2015)\citenamefont {Euv\'e},
  \citenamefont {Michel}, \citenamefont {Parentani},\ and\ \citenamefont
  {Rousseaux}}]{Euve2015}%
  \BibitemOpen
  \bibfield  {author} {\bibinfo {author} {\bibfnamefont {L.-P.}\ \bibnamefont
  {Euv\'e}}, \bibinfo {author} {\bibfnamefont {F.}~\bibnamefont {Michel}},
  \bibinfo {author} {\bibfnamefont {R.}~\bibnamefont {Parentani}}, \ and\
  \bibinfo {author} {\bibfnamefont {G.}~\bibnamefont {Rousseaux}},\ }\href
  {\doibase 10.1103/PhysRevD.91.024020} {\bibfield  {journal} {\bibinfo
  {journal} {Physical Review D}\ }\textbf {\bibinfo {volume} {91}},\ \bibinfo
  {pages} {024020} (\bibinfo {year} {2015})}\BibitemShut {NoStop}%
\bibitem [{\citenamefont {Philbin}\ \emph {et~al.}(2008)\citenamefont
  {Philbin}, \citenamefont {Kuklewicz}, \citenamefont {Robertson},
  \citenamefont {Hill}, \citenamefont {K\"onig},\ and\ \citenamefont
  {Leonhardt}}]{Philbin2008}%
  \BibitemOpen
  \bibfield  {author} {\bibinfo {author} {\bibfnamefont {T.~G.}\ \bibnamefont
  {Philbin}}, \bibinfo {author} {\bibfnamefont {C.}~\bibnamefont {Kuklewicz}},
  \bibinfo {author} {\bibfnamefont {S.}~\bibnamefont {Robertson}}, \bibinfo
  {author} {\bibfnamefont {S.}~\bibnamefont {Hill}}, \bibinfo {author}
  {\bibfnamefont {F.}~\bibnamefont {K\"onig}}, \ and\ \bibinfo {author}
  {\bibfnamefont {U.}~\bibnamefont {Leonhardt}},\ }\href {\doibase
  10.1126/science.1153625} {\bibfield  {journal} {\bibinfo  {journal}
  {Science}\ }\textbf {\bibinfo {volume} {319}},\ \bibinfo {pages} {1367}
  (\bibinfo {year} {2008})}\BibitemShut {NoStop}%
\bibitem [{\citenamefont {Steinhauer}(2016)}]{Steinhauer2016}%
  \BibitemOpen
  \bibfield  {author} {\bibinfo {author} {\bibfnamefont {J.}~\bibnamefont
  {Steinhauer}},\ }\href {http://dx.doi.org/10.1038/nphys3863} {\bibfield
  {journal} {\bibinfo  {journal} {Nature Physics}\ }\textbf {\bibinfo {volume}
  {12}},\ \bibinfo {pages} {959} (\bibinfo {year} {2016})}\BibitemShut
  {NoStop}%
\bibitem [{\citenamefont {Cropp}\ \emph {et~al.}(2016)\citenamefont {Cropp},
  \citenamefont {Liberati},\ and\ \citenamefont {Turcati}}]{Cropp2016}%
  \BibitemOpen
  \bibfield  {author} {\bibinfo {author} {\bibfnamefont {B.}~\bibnamefont
  {Cropp}}, \bibinfo {author} {\bibfnamefont {S.}~\bibnamefont {Liberati}}, \
  and\ \bibinfo {author} {\bibfnamefont {R.}~\bibnamefont {Turcati}},\ }\href
  {\doibase 10.1103/PhysRevD.94.063003} {\bibfield  {journal} {\bibinfo
  {journal} {Physical Review D}\ }\textbf {\bibinfo {volume} {94}},\ \bibinfo
  {pages} {063003} (\bibinfo {year} {2016})}\BibitemShut {NoStop}%
\bibitem [{\citenamefont {Dey}\ \emph {et~al.}(2016)\citenamefont {Dey},
  \citenamefont {Liberati},\ and\ \citenamefont {Turcati}}]{Dey2016}%
  \BibitemOpen
  \bibfield  {author} {\bibinfo {author} {\bibfnamefont {R.}~\bibnamefont
  {Dey}}, \bibinfo {author} {\bibfnamefont {S.}~\bibnamefont {Liberati}}, \
  and\ \bibinfo {author} {\bibfnamefont {R.}~\bibnamefont {Turcati}},\ }\href
  {\doibase 10.1103/PhysRevD.94.104068} {\bibfield  {journal} {\bibinfo
  {journal} {Physical Review D}\ }\textbf {\bibinfo {volume} {94}},\ \bibinfo
  {pages} {104068} (\bibinfo {year} {2016})}\BibitemShut {NoStop}%
\bibitem [{\citenamefont {Giacomelli}\ and\ \citenamefont
  {Liberati}(2017)}]{Giacomelli2017}%
  \BibitemOpen
  \bibfield  {author} {\bibinfo {author} {\bibfnamefont {L.}~\bibnamefont
  {Giacomelli}}\ and\ \bibinfo {author} {\bibfnamefont {S.}~\bibnamefont
  {Liberati}},\ }\href {\doibase 10.1103/PhysRevD.96.064014} {\bibfield
  {journal} {\bibinfo  {journal} {Physical Review D}\ }\textbf {\bibinfo
  {volume} {96}},\ \bibinfo {pages} {064014} (\bibinfo {year}
  {2017})}\BibitemShut {NoStop}%
\bibitem [{\citenamefont {Barcel\'o}\ \emph {et~al.}(2003)\citenamefont
  {Barcel\'o}, \citenamefont {Liberati},\ and\ \citenamefont
  {Visser}}]{Barcelo2003}%
  \BibitemOpen
  \bibfield  {author} {\bibinfo {author} {\bibfnamefont {C.}~\bibnamefont
  {Barcel\'o}}, \bibinfo {author} {\bibfnamefont {S.}~\bibnamefont {Liberati}},
  \ and\ \bibinfo {author} {\bibfnamefont {M.}~\bibnamefont {Visser}},\ }\href
  {\doibase 10.1142/S0218271803004092} {\bibfield  {journal} {\bibinfo
  {journal} {International Journal of Modern Physics D}\ }\textbf {\bibinfo
  {volume} {12}},\ \bibinfo {pages} {1641} (\bibinfo {year}
  {2003})}\BibitemShut {NoStop}%
\bibitem [{\citenamefont {Fagnocchi}\ \emph {et~al.}(2010)\citenamefont
  {Fagnocchi}, \citenamefont {Finazzi}, \citenamefont {Liberati}, \citenamefont
  {Kormos},\ and\ \citenamefont {Trombettoni}}]{Fagnocchi2010}%
  \BibitemOpen
  \bibfield  {author} {\bibinfo {author} {\bibfnamefont {S.}~\bibnamefont
  {Fagnocchi}}, \bibinfo {author} {\bibfnamefont {S.}~\bibnamefont {Finazzi}},
  \bibinfo {author} {\bibfnamefont {S.}~\bibnamefont {Liberati}}, \bibinfo
  {author} {\bibfnamefont {M.}~\bibnamefont {Kormos}}, \ and\ \bibinfo {author}
  {\bibfnamefont {A.}~\bibnamefont {Trombettoni}},\ }\href
  {http://stacks.iop.org/1367-2630/12/i=9/a=095012} {\bibfield  {journal}
  {\bibinfo  {journal} {New Journal of Physics}\ }\textbf {\bibinfo {volume}
  {12}},\ \bibinfo {pages} {095012} (\bibinfo {year} {2010})}\BibitemShut
  {NoStop}%
\bibitem [{\citenamefont {Fischer}\ and\ \citenamefont
  {Sch\"utzhold}(2004)}]{Fischer2004}%
  \BibitemOpen
  \bibfield  {author} {\bibinfo {author} {\bibfnamefont {U.~R.}\ \bibnamefont
  {Fischer}}\ and\ \bibinfo {author} {\bibfnamefont {R.}~\bibnamefont
  {Sch\"utzhold}},\ }\href {\doibase 10.1103/PhysRevA.70.063615} {\bibfield
  {journal} {\bibinfo  {journal} {Phys. Rev. A}\ }\textbf {\bibinfo {volume}
  {70}},\ \bibinfo {pages} {063615} (\bibinfo {year} {2004})}\BibitemShut
  {NoStop}%
\bibitem [{\citenamefont {Ch\"a}\ and\ \citenamefont
  {Fischer}(2017)}]{Cha2017}%
  \BibitemOpen
  \bibfield  {author} {\bibinfo {author} {\bibfnamefont {S.-Y.}\ \bibnamefont
  {Ch\"a}}\ and\ \bibinfo {author} {\bibfnamefont {U.~R.}\ \bibnamefont
  {Fischer}},\ }\href {\doibase 10.1103/PhysRevLett.118.130404} {\bibfield
  {journal} {\bibinfo  {journal} {Phys. Rev. Lett.}\ }\textbf {\bibinfo
  {volume} {118}},\ \bibinfo {pages} {130404} (\bibinfo {year}
  {2017})}\BibitemShut {NoStop}%
\bibitem [{\citenamefont {Bravo}\ \emph {et~al.}(2015)\citenamefont {Bravo},
  \citenamefont {Sab\'in},\ and\ \citenamefont {Fuentes}}]{Bravo2015}%
  \BibitemOpen
  \bibfield  {author} {\bibinfo {author} {\bibfnamefont {T.}~\bibnamefont
  {Bravo}}, \bibinfo {author} {\bibfnamefont {C.}~\bibnamefont {Sab\'in}}, \
  and\ \bibinfo {author} {\bibfnamefont {I.}~\bibnamefont {Fuentes}},\ }\href
  {https://epjquantumtechnology.springeropen.com/articles/10.1140/epjqt16}
  {\bibfield  {journal} {\bibinfo  {journal} {EPJ Quantum Technology}\ }\textbf
  {\bibinfo {volume} {2}},\ \bibinfo {pages} {3} (\bibinfo {year}
  {2015})}\BibitemShut {NoStop}%
\bibitem [{\citenamefont {Abbott}\ \emph
  {et~al.}(2016{\natexlab{a}})\citenamefont {Abbott}, \citenamefont {Abbott},
  \citenamefont {Abbott}, \citenamefont {Abernathy}, \citenamefont {Acernese},
  \citenamefont {Ackley}, \citenamefont {Adams}, \citenamefont {Adams},
  \citenamefont {Addesso}, \citenamefont {Adhikari}, \citenamefont {Adya},
  \citenamefont {Affeldt}, \citenamefont {Agathos}, \citenamefont {Agatsuma},
  \citenamefont {Aggarwal} \emph {et~al.}}]{Abbott2016}%
  \BibitemOpen
  \bibfield  {author} {\bibinfo {author} {\bibfnamefont {B.~P.}\ \bibnamefont
  {Abbott}}, \bibinfo {author} {\bibfnamefont {R.}~\bibnamefont {Abbott}},
  \bibinfo {author} {\bibfnamefont {T.~D.}\ \bibnamefont {Abbott}}, \bibinfo
  {author} {\bibfnamefont {M.~R.}\ \bibnamefont {Abernathy}}, \bibinfo {author}
  {\bibfnamefont {F.}~\bibnamefont {Acernese}}, \bibinfo {author}
  {\bibfnamefont {K.}~\bibnamefont {Ackley}}, \bibinfo {author} {\bibfnamefont
  {C.}~\bibnamefont {Adams}}, \bibinfo {author} {\bibfnamefont
  {T.}~\bibnamefont {Adams}}, \bibinfo {author} {\bibfnamefont
  {P.}~\bibnamefont {Addesso}}, \bibinfo {author} {\bibfnamefont {R.~X.}\
  \bibnamefont {Adhikari}}, \bibinfo {author} {\bibfnamefont {V.~B.}\
  \bibnamefont {Adya}}, \bibinfo {author} {\bibfnamefont {C.}~\bibnamefont
  {Affeldt}}, \bibinfo {author} {\bibfnamefont {M.}~\bibnamefont {Agathos}},
  \bibinfo {author} {\bibfnamefont {K.}~\bibnamefont {Agatsuma}}, \bibinfo
  {author} {\bibfnamefont {N.}~\bibnamefont {Aggarwal}},  \emph {et~al.}
  (\bibinfo {collaboration} {LIGO Scientific Collaboration and Virgo
  Collaboration}),\ }\href {\doibase 10.1103/PhysRevLett.116.061102} {\bibfield
   {journal} {\bibinfo  {journal} {Physical Review Letters}\ }\textbf {\bibinfo
  {volume} {116}},\ \bibinfo {pages} {061102} (\bibinfo {year}
  {2016}{\natexlab{a}})}\BibitemShut {NoStop}%
\bibitem [{\citenamefont {Abbott}\ \emph
  {et~al.}(2016{\natexlab{b}})\citenamefont {Abbott}, \citenamefont {Abbott},
  \citenamefont {Abbott}, \citenamefont {Abernathy}, \citenamefont {Acernese},
  \citenamefont {Ackley}, \citenamefont {Adams}, \citenamefont {Adams},
  \citenamefont {Addesso}, \citenamefont {Adhikari}, \citenamefont {Adya},
  \citenamefont {Affeldt}, \citenamefont {Agathos}, \citenamefont {Agatsuma},
  \citenamefont {Aggarwal} \emph {et~al.}}]{Abbott2016a}%
  \BibitemOpen
  \bibfield  {author} {\bibinfo {author} {\bibfnamefont {B.~P.}\ \bibnamefont
  {Abbott}}, \bibinfo {author} {\bibfnamefont {R.}~\bibnamefont {Abbott}},
  \bibinfo {author} {\bibfnamefont {T.~D.}\ \bibnamefont {Abbott}}, \bibinfo
  {author} {\bibfnamefont {M.~R.}\ \bibnamefont {Abernathy}}, \bibinfo {author}
  {\bibfnamefont {F.}~\bibnamefont {Acernese}}, \bibinfo {author}
  {\bibfnamefont {K.}~\bibnamefont {Ackley}}, \bibinfo {author} {\bibfnamefont
  {C.}~\bibnamefont {Adams}}, \bibinfo {author} {\bibfnamefont
  {T.}~\bibnamefont {Adams}}, \bibinfo {author} {\bibfnamefont
  {P.}~\bibnamefont {Addesso}}, \bibinfo {author} {\bibfnamefont {R.~X.}\
  \bibnamefont {Adhikari}}, \bibinfo {author} {\bibfnamefont {V.~B.}\
  \bibnamefont {Adya}}, \bibinfo {author} {\bibfnamefont {C.}~\bibnamefont
  {Affeldt}}, \bibinfo {author} {\bibfnamefont {M.}~\bibnamefont {Agathos}},
  \bibinfo {author} {\bibfnamefont {K.}~\bibnamefont {Agatsuma}}, \bibinfo
  {author} {\bibfnamefont {N.}~\bibnamefont {Aggarwal}},  \emph {et~al.}
  (\bibinfo {collaboration} {LIGO Scientific Collaboration and Virgo
  Collaboration}),\ }\href {\doibase 10.1103/PhysRevLett.116.241103} {\bibfield
   {journal} {\bibinfo  {journal} {Physical Review Letters}\ }\textbf {\bibinfo
  {volume} {116}},\ \bibinfo {pages} {241103} (\bibinfo {year}
  {2016}{\natexlab{b}})}\BibitemShut {NoStop}%
\bibitem [{\citenamefont {Abbott}\ \emph
  {et~al.}(2017{\natexlab{a}})\citenamefont {Abbott}, \citenamefont {Abbott},
  \citenamefont {Abbott}, \citenamefont {Acernese}, \citenamefont {Ackley},
  \citenamefont {Adams}, \citenamefont {Adams}, \citenamefont {Addesso},
  \citenamefont {Adhikari}, \citenamefont {Adya}, \citenamefont {Affeldt},
  \citenamefont {Afrough}, \citenamefont {Agarwal}, \citenamefont {Agathos},
  \citenamefont {Agatsuma} \emph {et~al.}}]{Abbott2017}%
  \BibitemOpen
  \bibfield  {author} {\bibinfo {author} {\bibfnamefont {B.~P.}\ \bibnamefont
  {Abbott}}, \bibinfo {author} {\bibfnamefont {R.}~\bibnamefont {Abbott}},
  \bibinfo {author} {\bibfnamefont {T.~D.}\ \bibnamefont {Abbott}}, \bibinfo
  {author} {\bibfnamefont {F.}~\bibnamefont {Acernese}}, \bibinfo {author}
  {\bibfnamefont {K.}~\bibnamefont {Ackley}}, \bibinfo {author} {\bibfnamefont
  {C.}~\bibnamefont {Adams}}, \bibinfo {author} {\bibfnamefont
  {T.}~\bibnamefont {Adams}}, \bibinfo {author} {\bibfnamefont
  {P.}~\bibnamefont {Addesso}}, \bibinfo {author} {\bibfnamefont {R.~X.}\
  \bibnamefont {Adhikari}}, \bibinfo {author} {\bibfnamefont {V.~B.}\
  \bibnamefont {Adya}}, \bibinfo {author} {\bibfnamefont {C.}~\bibnamefont
  {Affeldt}}, \bibinfo {author} {\bibfnamefont {M.}~\bibnamefont {Afrough}},
  \bibinfo {author} {\bibfnamefont {B.}~\bibnamefont {Agarwal}}, \bibinfo
  {author} {\bibfnamefont {M.}~\bibnamefont {Agathos}}, \bibinfo {author}
  {\bibfnamefont {K.}~\bibnamefont {Agatsuma}},  \emph {et~al.} (\bibinfo
  {collaboration} {LIGO Scientific and Virgo Collaboration}),\ }\href {\doibase
  10.1103/PhysRevLett.118.221101} {\bibfield  {journal} {\bibinfo  {journal}
  {Physical Review Letters}\ }\textbf {\bibinfo {volume} {118}},\ \bibinfo
  {pages} {221101} (\bibinfo {year} {2017}{\natexlab{a}})}\BibitemShut
  {NoStop}%
\bibitem [{\citenamefont {Abbott}\ \emph
  {et~al.}(2017{\natexlab{b}})\citenamefont {Abbott}, \citenamefont {Abbott},
  \citenamefont {Abbott}, \citenamefont {Acernese}, \citenamefont {Ackley},
  \citenamefont {Adams}, \citenamefont {Adams}, \citenamefont {Addesso},
  \citenamefont {Adhikari}, \citenamefont {Adya}, \citenamefont {Affeldt},
  \citenamefont {Afrough}, \citenamefont {Agarwal}, \citenamefont {Agathos},
  \citenamefont {Agatsuma} \emph {et~al.}}]{Abbott2017a}%
  \BibitemOpen
  \bibfield  {author} {\bibinfo {author} {\bibfnamefont {B.~P.}\ \bibnamefont
  {Abbott}}, \bibinfo {author} {\bibfnamefont {R.}~\bibnamefont {Abbott}},
  \bibinfo {author} {\bibfnamefont {T.~D.}\ \bibnamefont {Abbott}}, \bibinfo
  {author} {\bibfnamefont {F.}~\bibnamefont {Acernese}}, \bibinfo {author}
  {\bibfnamefont {K.}~\bibnamefont {Ackley}}, \bibinfo {author} {\bibfnamefont
  {C.}~\bibnamefont {Adams}}, \bibinfo {author} {\bibfnamefont
  {T.}~\bibnamefont {Adams}}, \bibinfo {author} {\bibfnamefont
  {P.}~\bibnamefont {Addesso}}, \bibinfo {author} {\bibfnamefont {R.~X.}\
  \bibnamefont {Adhikari}}, \bibinfo {author} {\bibfnamefont {V.~B.}\
  \bibnamefont {Adya}}, \bibinfo {author} {\bibfnamefont {C.}~\bibnamefont
  {Affeldt}}, \bibinfo {author} {\bibfnamefont {M.}~\bibnamefont {Afrough}},
  \bibinfo {author} {\bibfnamefont {B.}~\bibnamefont {Agarwal}}, \bibinfo
  {author} {\bibfnamefont {M.}~\bibnamefont {Agathos}}, \bibinfo {author}
  {\bibfnamefont {K.}~\bibnamefont {Agatsuma}},  \emph {et~al.} (\bibinfo
  {collaboration} {LIGO Scientific Collaboration and Virgo Collaboration}),\
  }\href {\doibase 10.1103/PhysRevLett.119.141101} {\bibfield  {journal}
  {\bibinfo  {journal} {Physical Review Letters}\ }\textbf {\bibinfo {volume}
  {119}},\ \bibinfo {pages} {141101} (\bibinfo {year}
  {2017}{\natexlab{b}})}\BibitemShut {NoStop}%
\bibitem [{\citenamefont {Abbott}\ \emph
  {et~al.}(2017{\natexlab{c}})\citenamefont {Abbott}, \citenamefont {Abbott},
  \citenamefont {Abbott}, \citenamefont {Acernese}, \citenamefont {Ackley},
  \citenamefont {Adams}, \citenamefont {Adams}, \citenamefont {Addesso},
  \citenamefont {Adhikari}, \citenamefont {Adya}, \citenamefont {Affeldt},
  \citenamefont {Afrough}, \citenamefont {Agarwal}, \citenamefont {Agathos},
  \citenamefont {Agatsuma} \emph {et~al.}}]{Abbott2017b}%
  \BibitemOpen
  \bibfield  {author} {\bibinfo {author} {\bibfnamefont {B.~P.}\ \bibnamefont
  {Abbott}}, \bibinfo {author} {\bibfnamefont {R.}~\bibnamefont {Abbott}},
  \bibinfo {author} {\bibfnamefont {T.~D.}\ \bibnamefont {Abbott}}, \bibinfo
  {author} {\bibfnamefont {F.}~\bibnamefont {Acernese}}, \bibinfo {author}
  {\bibfnamefont {K.}~\bibnamefont {Ackley}}, \bibinfo {author} {\bibfnamefont
  {C.}~\bibnamefont {Adams}}, \bibinfo {author} {\bibfnamefont
  {T.}~\bibnamefont {Adams}}, \bibinfo {author} {\bibfnamefont
  {P.}~\bibnamefont {Addesso}}, \bibinfo {author} {\bibfnamefont {R.~X.}\
  \bibnamefont {Adhikari}}, \bibinfo {author} {\bibfnamefont {V.~B.}\
  \bibnamefont {Adya}}, \bibinfo {author} {\bibfnamefont {C.}~\bibnamefont
  {Affeldt}}, \bibinfo {author} {\bibfnamefont {M.}~\bibnamefont {Afrough}},
  \bibinfo {author} {\bibfnamefont {B.}~\bibnamefont {Agarwal}}, \bibinfo
  {author} {\bibfnamefont {M.}~\bibnamefont {Agathos}}, \bibinfo {author}
  {\bibfnamefont {K.}~\bibnamefont {Agatsuma}},  \emph {et~al.} (\bibinfo
  {collaboration} {LIGO Scientific Collaboration and Virgo Collaboration}),\
  }\href {\doibase 10.1103/PhysRevLett.119.161101} {\bibfield  {journal}
  {\bibinfo  {journal} {Physical Review Letters}\ }\textbf {\bibinfo {volume}
  {119}},\ \bibinfo {pages} {161101} (\bibinfo {year}
  {2017}{\natexlab{c}})}\BibitemShut {NoStop}%
\bibitem [{\citenamefont {Fernandez-Corbaton}\ \emph
  {et~al.}(2015)\citenamefont {Fernandez-Corbaton}, \citenamefont {Cirio},
  \citenamefont {Büse}, \citenamefont {Lamata}, \citenamefont {Solano},\ and\
  \citenamefont {Molina-Terriza}}]{Fernandez-Corbaton2015}%
  \BibitemOpen
  \bibfield  {author} {\bibinfo {author} {\bibfnamefont {I.}~\bibnamefont
  {Fernandez-Corbaton}}, \bibinfo {author} {\bibfnamefont {M.}~\bibnamefont
  {Cirio}}, \bibinfo {author} {\bibfnamefont {A.}~\bibnamefont {Büse}},
  \bibinfo {author} {\bibfnamefont {L.}~\bibnamefont {Lamata}}, \bibinfo
  {author} {\bibfnamefont {E.}~\bibnamefont {Solano}}, \ and\ \bibinfo {author}
  {\bibfnamefont {G.}~\bibnamefont {Molina-Terriza}},\ }\href
  {http://dx.doi.org/10.1038/srep11538} {\bibfield  {journal} {\bibinfo
  {journal} {Scientific Reports}\ }\textbf {\bibinfo {volume} {5}},\ \bibinfo
  {pages} {11538} (\bibinfo {year} {2015})}\BibitemShut {NoStop}%
\bibitem [{\citenamefont {Jones}\ \emph {et~al.}(2017)\citenamefont {Jones},
  \citenamefont {McDougall},\ and\ \citenamefont {Singleton}}]{Jones2017}%
  \BibitemOpen
  \bibfield  {author} {\bibinfo {author} {\bibfnamefont {P.}~\bibnamefont
  {Jones}}, \bibinfo {author} {\bibfnamefont {P.}~\bibnamefont {McDougall}}, \
  and\ \bibinfo {author} {\bibfnamefont {D.}~\bibnamefont {Singleton}},\ }\href
  {\doibase 10.1103/PhysRevD.95.065010} {\bibfield  {journal} {\bibinfo
  {journal} {Physical Review D}\ }\textbf {\bibinfo {volume} {95}},\ \bibinfo
  {pages} {065010} (\bibinfo {year} {2017})}\BibitemShut {NoStop}%
\bibitem [{\citenamefont {Jaekel}\ \emph {et~al.}(2006)\citenamefont {Jaekel},
  \citenamefont {Lamine}, \citenamefont {Lambrecht}, \citenamefont {Reynaud},\
  and\ \citenamefont {Neto}}]{Jaekel2006}%
  \BibitemOpen
  \bibfield  {author} {\bibinfo {author} {\bibfnamefont {M.-T.}\ \bibnamefont
  {Jaekel}}, \bibinfo {author} {\bibfnamefont {B.}~\bibnamefont {Lamine}},
  \bibinfo {author} {\bibfnamefont {A.}~\bibnamefont {Lambrecht}}, \bibinfo
  {author} {\bibfnamefont {S.}~\bibnamefont {Reynaud}}, \ and\ \bibinfo
  {author} {\bibfnamefont {P.~M.}\ \bibnamefont {Neto}},\ }in\ \href
  {https://hal.archives-ouvertes.fr/hal-00288533} {\emph {\bibinfo {booktitle}
  {Proceedings Leiden 2006}}}\ (\bibinfo  {publisher} {World Scientific},\
  \bibinfo {address} {Leiden, Netherlands},\ \bibinfo {year} {2006})\ p.\
  \bibinfo {pages} {125},\ \bibinfo {note} {7 pages}\BibitemShut {NoStop}%
\bibitem [{\citenamefont {Sab\'in}\ \emph {et~al.}(2014)\citenamefont
  {Sab\'in}, \citenamefont {Bruschi}, \citenamefont {Ahmadi},\ and\
  \citenamefont {Fuentes}}]{Sabin2014}%
  \BibitemOpen
  \bibfield  {author} {\bibinfo {author} {\bibfnamefont {C.}~\bibnamefont
  {Sab\'in}}, \bibinfo {author} {\bibfnamefont {D.~E.}\ \bibnamefont
  {Bruschi}}, \bibinfo {author} {\bibfnamefont {M.}~\bibnamefont {Ahmadi}}, \
  and\ \bibinfo {author} {\bibfnamefont {I.}~\bibnamefont {Fuentes}},\ }\href
  {http://stacks.iop.org/1367-2630/16/i=8/a=085003} {\bibfield  {journal}
  {\bibinfo  {journal} {New Journal of Physics}\ }\textbf {\bibinfo {volume}
  {16}},\ \bibinfo {pages} {085003} (\bibinfo {year} {2014})}\BibitemShut
  {NoStop}%
\bibitem [{\citenamefont {Sab\'in}\ \emph {et~al.}(2016)\citenamefont
  {Sab\'in}, \citenamefont {Kohlrus}, \citenamefont {Bruschi},\ and\
  \citenamefont {Fuentes}}]{Sabin2016}%
  \BibitemOpen
  \bibfield  {author} {\bibinfo {author} {\bibfnamefont {C.}~\bibnamefont
  {Sab\'in}}, \bibinfo {author} {\bibfnamefont {J.}~\bibnamefont {Kohlrus}},
  \bibinfo {author} {\bibfnamefont {D.~E.}\ \bibnamefont {Bruschi}}, \ and\
  \bibinfo {author} {\bibfnamefont {I.}~\bibnamefont {Fuentes}},\ }\href
  {http://dx.doi.org/10.1140/epjqt/s40507-016-0046-4} {\bibfield  {journal}
  {\bibinfo  {journal} {EPJ Quantum Technology}\ }\textbf {\bibinfo {volume}
  {3}} (\bibinfo {year} {2016})}\BibitemShut {NoStop}%
\bibitem [{\citenamefont {Howl}\ \emph {et~al.}(2016)\citenamefont {Howl},
  \citenamefont {Hackerm\"uller}, \citenamefont {Bruschi},\ and\ \citenamefont
  {Fuentes}}]{Howl2016a}%
  \BibitemOpen
  \bibfield  {author} {\bibinfo {author} {\bibfnamefont {R.}~\bibnamefont
  {Howl}}, \bibinfo {author} {\bibfnamefont {L.}~\bibnamefont
  {Hackerm\"uller}}, \bibinfo {author} {\bibfnamefont {D.~E.}\ \bibnamefont
  {Bruschi}}, \ and\ \bibinfo {author} {\bibfnamefont {I.}~\bibnamefont
  {Fuentes}},\ }\href {https://arxiv.org/abs/1607.06666} {\bibfield  {journal}
  {\bibinfo  {journal} {arXiv:1607.06666v2}\ } (\bibinfo {year}
  {2016})}\BibitemShut {NoStop}%
\bibitem [{\citenamefont {Moore}(1970)}]{Moore1970}%
  \BibitemOpen
  \bibfield  {author} {\bibinfo {author} {\bibfnamefont {G.~T.}\ \bibnamefont
  {Moore}},\ }\href {\doibase 10.1063/1.1665432} {\bibfield  {journal}
  {\bibinfo  {journal} {Journal of Mathematical Physics}\ }\textbf {\bibinfo
  {volume} {11}},\ \bibinfo {pages} {2679} (\bibinfo {year}
  {1970})}\BibitemShut {NoStop}%
\bibitem [{\citenamefont {Fulling}\ and\ \citenamefont
  {Davies}(1976)}]{Fulling1976}%
  \BibitemOpen
  \bibfield  {author} {\bibinfo {author} {\bibfnamefont {S.~A.}\ \bibnamefont
  {Fulling}}\ and\ \bibinfo {author} {\bibfnamefont {P.~C.~W.}\ \bibnamefont
  {Davies}},\ }\href {\doibase 10.1098/rspa.1976.0045} {\bibfield  {journal}
  {\bibinfo  {journal} {Proceedings of the Royal Society of London A:
  Mathematical, Physical and Engineering Sciences}\ }\textbf {\bibinfo {volume}
  {348}},\ \bibinfo {pages} {393} (\bibinfo {year} {1976})}\BibitemShut
  {NoStop}%
\bibitem [{\citenamefont {Howl}\ \emph {et~al.}(2018)\citenamefont {Howl},
  \citenamefont {Sab\'in}, \citenamefont {Hackerm\"uller},\ and\ \citenamefont
  {Fuentes}}]{Howl2016}%
  \BibitemOpen
  \bibfield  {author} {\bibinfo {author} {\bibfnamefont {R.}~\bibnamefont
  {Howl}}, \bibinfo {author} {\bibfnamefont {C.}~\bibnamefont {Sab\'in}},
  \bibinfo {author} {\bibfnamefont {L.}~\bibnamefont {Hackerm\"uller}}, \ and\
  \bibinfo {author} {\bibfnamefont {I.}~\bibnamefont {Fuentes}},\ }\href
  {http://stacks.iop.org/0953-4075/51/i=1/a=015303} {\bibfield  {journal}
  {\bibinfo  {journal} {Journal of Physics B: Atomic, Molecular and Optical
  Physics}\ }\textbf {\bibinfo {volume} {51}},\ \bibinfo {pages} {015303}
  (\bibinfo {year} {2018})}\BibitemShut {NoStop}%
\bibitem [{\citenamefont {Braunstein}\ and\ \citenamefont
  {Caves}(1994)}]{Braunstein1994}%
  \BibitemOpen
  \bibfield  {author} {\bibinfo {author} {\bibfnamefont {S.~L.}\ \bibnamefont
  {Braunstein}}\ and\ \bibinfo {author} {\bibfnamefont {C.~M.}\ \bibnamefont
  {Caves}},\ }\href {\doibase 10.1103/PhysRevLett.72.3439} {\bibfield
  {journal} {\bibinfo  {journal} {Physical Review Letters}\ }\textbf {\bibinfo
  {volume} {72}},\ \bibinfo {pages} {3439} (\bibinfo {year}
  {1994})}\BibitemShut {NoStop}%
\bibitem [{\citenamefont {Misner}\ \emph {et~al.}(1973)\citenamefont {Misner},
  \citenamefont {Thorne},\ and\ \citenamefont {Wheeler}}]{MTW}%
  \BibitemOpen
  \bibfield  {author} {\bibinfo {author} {\bibfnamefont {C.~W.}\ \bibnamefont
  {Misner}}, \bibinfo {author} {\bibfnamefont {K.~S.}\ \bibnamefont {Thorne}},
  \ and\ \bibinfo {author} {\bibfnamefont {J.~A.}\ \bibnamefont {Wheeler}},\
  }\href@noop {} {\emph {\bibinfo {title} {Gravitation}}}\ (\bibinfo
  {publisher} {W. H. Freeman and Company},\ \bibinfo {year} {1973})\BibitemShut
  {NoStop}%
\bibitem [{\citenamefont {Maggiore}(2007)}]{MaggioreGW}%
  \BibitemOpen
  \bibfield  {author} {\bibinfo {author} {\bibfnamefont {M.}~\bibnamefont
  {Maggiore}},\ }\href {http://www.oupcanada.com/catalog/9780198570745.html}
  {\emph {\bibinfo {title} {Gravitational Waves}}}\ (\bibinfo  {publisher}
  {Oxford University Press Canada},\ \bibinfo {year} {2007})\BibitemShut
  {NoStop}%
\bibitem [{\citenamefont {Visser}\ and\ \citenamefont
  {Molina-Paris}(2010)}]{Visser2010}%
  \BibitemOpen
  \bibfield  {author} {\bibinfo {author} {\bibfnamefont {M.}~\bibnamefont
  {Visser}}\ and\ \bibinfo {author} {\bibfnamefont {C.}~\bibnamefont
  {Molina-Paris}},\ }\href {http://stacks.iop.org/1367-2630/12/i=9/a=095014}
  {\bibfield  {journal} {\bibinfo  {journal} {New Journal of Physics}\ }\textbf
  {\bibinfo {volume} {12}},\ \bibinfo {pages} {095014} (\bibinfo {year}
  {2010})}\BibitemShut {NoStop}%
\bibitem [{\citenamefont {Bruschi}\ \emph {et~al.}(2014)\citenamefont
  {Bruschi}, \citenamefont {Sab\'in}, \citenamefont {White}, \citenamefont
  {Baccetti}, \citenamefont {Oi},\ and\ \citenamefont {Fuentes}}]{Bruschi2014}%
  \BibitemOpen
  \bibfield  {author} {\bibinfo {author} {\bibfnamefont {D.~E.}\ \bibnamefont
  {Bruschi}}, \bibinfo {author} {\bibfnamefont {C.}~\bibnamefont {Sab\'in}},
  \bibinfo {author} {\bibfnamefont {A.}~\bibnamefont {White}}, \bibinfo
  {author} {\bibfnamefont {V.}~\bibnamefont {Baccetti}}, \bibinfo {author}
  {\bibfnamefont {D.~K.~L.}\ \bibnamefont {Oi}}, \ and\ \bibinfo {author}
  {\bibfnamefont {I.}~\bibnamefont {Fuentes}},\ }\href
  {http://stacks.iop.org/1367-2630/16/i=5/a=053041} {\bibfield  {journal}
  {\bibinfo  {journal} {New Journal of Physics}\ }\textbf {\bibinfo {volume}
  {16}},\ \bibinfo {pages} {053041} (\bibinfo {year} {2014})}\BibitemShut
  {NoStop}%
\bibitem [{\citenamefont {Schneider}\ \emph {et~al.}(2012)\citenamefont
  {Schneider}, \citenamefont {Hackerm\"uller}, \citenamefont {Ronzheimer},
  \citenamefont {Will}, \citenamefont {Braun}, \citenamefont {Best},
  \citenamefont {Bloch}, \citenamefont {Demler}, \citenamefont {Mandt},
  \citenamefont {Rasch},\ and\ \citenamefont {Rosch}}]{Schneider2012}%
  \BibitemOpen
  \bibfield  {author} {\bibinfo {author} {\bibfnamefont {U.}~\bibnamefont
  {Schneider}}, \bibinfo {author} {\bibfnamefont {L.}~\bibnamefont
  {Hackerm\"uller}}, \bibinfo {author} {\bibfnamefont {J.~P.}\ \bibnamefont
  {Ronzheimer}}, \bibinfo {author} {\bibfnamefont {S.}~\bibnamefont {Will}},
  \bibinfo {author} {\bibfnamefont {S.}~\bibnamefont {Braun}}, \bibinfo
  {author} {\bibfnamefont {T.}~\bibnamefont {Best}}, \bibinfo {author}
  {\bibfnamefont {I.}~\bibnamefont {Bloch}}, \bibinfo {author} {\bibfnamefont
  {E.}~\bibnamefont {Demler}}, \bibinfo {author} {\bibfnamefont
  {S.}~\bibnamefont {Mandt}}, \bibinfo {author} {\bibfnamefont
  {D.}~\bibnamefont {Rasch}}, \ and\ \bibinfo {author} {\bibfnamefont
  {A.}~\bibnamefont {Rosch}},\ }\href {http://dx.doi.org/10.1038/nphys2205}
  {\bibfield  {journal} {\bibinfo  {journal} {Nature Physics}\ }\textbf
  {\bibinfo {volume} {8}},\ \bibinfo {pages} {213} (\bibinfo {year}
  {2012})}\BibitemShut {NoStop}%
\bibitem [{\citenamefont {Leonhardt}(2002)}]{Leonhardt2002}%
  \BibitemOpen
  \bibfield  {author} {\bibinfo {author} {\bibfnamefont {U.}~\bibnamefont
  {Leonhardt}},\ }\href {http://dx.doi.org/10.1038/415406a} {\bibfield
  {journal} {\bibinfo  {journal} {Nature}\ }\textbf {\bibinfo {volume} {415}},\
  \bibinfo {pages} {406} (\bibinfo {year} {2002})}\BibitemShut {NoStop}%
\bibitem [{\citenamefont {Richarte}\ \emph {et~al.}(2017)\citenamefont
  {Richarte}, \citenamefont {Salako}, \citenamefont
  {Gra\ifmmode~\mbox{\c{c}}\else \c{c}\fi{}a}, \citenamefont {Moradpour},\ and\
  \citenamefont {\"Ovg\"un}}]{Richarte2017}%
  \BibitemOpen
  \bibfield  {author} {\bibinfo {author} {\bibfnamefont {M.~G.}\ \bibnamefont
  {Richarte}}, \bibinfo {author} {\bibfnamefont {I.~G.}\ \bibnamefont
  {Salako}}, \bibinfo {author} {\bibfnamefont {J.~P.~M.}\ \bibnamefont
  {Gra\ifmmode~\mbox{\c{c}}\else \c{c}\fi{}a}}, \bibinfo {author}
  {\bibfnamefont {H.}~\bibnamefont {Moradpour}}, \ and\ \bibinfo {author}
  {\bibfnamefont {A.}~\bibnamefont {\"Ovg\"un}},\ }\href {\doibase
  10.1103/PhysRevD.96.084022} {\bibfield  {journal} {\bibinfo  {journal}
  {Physical Review D}\ }\textbf {\bibinfo {volume} {96}},\ \bibinfo {pages}
  {084022} (\bibinfo {year} {2017})}\BibitemShut {NoStop}%
\bibitem [{\citenamefont {Aasi}\ \emph
  {et~al.}(2014{\natexlab{a}})\citenamefont {Aasi}, \citenamefont {Abbott},
  \citenamefont {Abbott}, \citenamefont {Abbott}, \citenamefont {Abernathy},
  \citenamefont {Accadia}, \citenamefont {Acernese}, \citenamefont {Ackley},
  \citenamefont {Adams}, \citenamefont {Adams}, \citenamefont {Addesso},
  \citenamefont {Adhikari}, \citenamefont {Affeldt}, \citenamefont {Agathos},
  \citenamefont {Aggarwal} \emph {et~al.}}]{Aasi2014a}%
  \BibitemOpen
  \bibfield  {author} {\bibinfo {author} {\bibfnamefont {J.}~\bibnamefont
  {Aasi}}, \bibinfo {author} {\bibfnamefont {B.~P.}\ \bibnamefont {Abbott}},
  \bibinfo {author} {\bibfnamefont {R.}~\bibnamefont {Abbott}}, \bibinfo
  {author} {\bibfnamefont {T.}~\bibnamefont {Abbott}}, \bibinfo {author}
  {\bibfnamefont {M.~R.}\ \bibnamefont {Abernathy}}, \bibinfo {author}
  {\bibfnamefont {T.}~\bibnamefont {Accadia}}, \bibinfo {author} {\bibfnamefont
  {F.}~\bibnamefont {Acernese}}, \bibinfo {author} {\bibfnamefont
  {K.}~\bibnamefont {Ackley}}, \bibinfo {author} {\bibfnamefont
  {C.}~\bibnamefont {Adams}}, \bibinfo {author} {\bibfnamefont
  {T.}~\bibnamefont {Adams}}, \bibinfo {author} {\bibfnamefont
  {P.}~\bibnamefont {Addesso}}, \bibinfo {author} {\bibfnamefont {R.~X.}\
  \bibnamefont {Adhikari}}, \bibinfo {author} {\bibfnamefont {C.}~\bibnamefont
  {Affeldt}}, \bibinfo {author} {\bibfnamefont {M.}~\bibnamefont {Agathos}},
  \bibinfo {author} {\bibfnamefont {N.}~\bibnamefont {Aggarwal}},  \emph
  {et~al.} (\bibinfo {collaboration} {LIGO Scientific Collaboration and Virgo
  Collaboration}),\ }\href {\doibase 10.1103/PhysRevD.90.062010} {\bibfield
  {journal} {\bibinfo  {journal} {Physical Review D}\ }\textbf {\bibinfo
  {volume} {90}},\ \bibinfo {pages} {062010} (\bibinfo {year}
  {2014}{\natexlab{a}})}\BibitemShut {NoStop}%
\bibitem [{\citenamefont {Palomba}(2012)}]{Palomba2012}%
  \BibitemOpen
  \bibfield  {author} {\bibinfo {author} {\bibfnamefont {C.}~\bibnamefont
  {Palomba}},\ }\href {https://arxiv.org/abs/1201.3176} {\bibfield  {journal}
  {\bibinfo  {journal} {arXiv:1201.3176}\ } (\bibinfo {year}
  {2012})}\BibitemShut {NoStop}%
\bibitem [{\citenamefont {Aasi}\ \emph
  {et~al.}(2014{\natexlab{b}})\citenamefont {Aasi}, \citenamefont {Abbott},
  \citenamefont {Abbott}, \citenamefont {Abbott}, \citenamefont {Abernathy},
  \citenamefont {Acernese}, \citenamefont {Ackley}, \citenamefont {Adams},
  \citenamefont {Adams}, \citenamefont {Addesso}, \citenamefont {Adhikari},
  \citenamefont {Affeldt}, \citenamefont {Agathos}, \citenamefont {Aggarwal},
  \citenamefont {Aguiar} \emph {et~al.}}]{Aasi2014}%
  \BibitemOpen
  \bibfield  {author} {\bibinfo {author} {\bibfnamefont {J.}~\bibnamefont
  {Aasi}}, \bibinfo {author} {\bibfnamefont {B.~P.}\ \bibnamefont {Abbott}},
  \bibinfo {author} {\bibfnamefont {R.}~\bibnamefont {Abbott}}, \bibinfo
  {author} {\bibfnamefont {T.}~\bibnamefont {Abbott}}, \bibinfo {author}
  {\bibfnamefont {M.~R.}\ \bibnamefont {Abernathy}}, \bibinfo {author}
  {\bibfnamefont {F.}~\bibnamefont {Acernese}}, \bibinfo {author}
  {\bibfnamefont {K.}~\bibnamefont {Ackley}}, \bibinfo {author} {\bibfnamefont
  {C.}~\bibnamefont {Adams}}, \bibinfo {author} {\bibfnamefont
  {T.}~\bibnamefont {Adams}}, \bibinfo {author} {\bibfnamefont
  {P.}~\bibnamefont {Addesso}}, \bibinfo {author} {\bibfnamefont {R.~X.}\
  \bibnamefont {Adhikari}}, \bibinfo {author} {\bibfnamefont {C.}~\bibnamefont
  {Affeldt}}, \bibinfo {author} {\bibfnamefont {M.}~\bibnamefont {Agathos}},
  \bibinfo {author} {\bibfnamefont {N.}~\bibnamefont {Aggarwal}}, \bibinfo
  {author} {\bibfnamefont {O.~D.}\ \bibnamefont {Aguiar}},  \emph {et~al.}
  (\bibinfo {collaboration} {The LIGO Scientific Collaboration and the Virgo
  Collaboration}),\ }\href {\doibase 10.1103/PhysRevD.89.102006} {\bibfield
  {journal} {\bibinfo  {journal} {Physical Review D}\ }\textbf {\bibinfo
  {volume} {89}},\ \bibinfo {pages} {102006} (\bibinfo {year}
  {2014}{\natexlab{b}})}\BibitemShut {NoStop}%
\bibitem [{\citenamefont {Barcel\'o}\ \emph {et~al.}(2005)\citenamefont
  {Barcel\'o}, \citenamefont {Liberati},\ and\ \citenamefont
  {Visser}}]{Barcelo2005}%
  \BibitemOpen
  \bibfield  {author} {\bibinfo {author} {\bibfnamefont {C.}~\bibnamefont
  {Barcel\'o}}, \bibinfo {author} {\bibfnamefont {S.}~\bibnamefont {Liberati}},
  \ and\ \bibinfo {author} {\bibfnamefont {M.}~\bibnamefont {Visser}},\ }\href
  {\doibase 10.1007/lrr-2005-12} {\bibfield  {journal} {\bibinfo  {journal}
  {Living Reviews in Relativity}\ }\textbf {\bibinfo {volume} {8}} (\bibinfo
  {year} {2005}),\ 10.1007/lrr-2005-12}\BibitemShut {NoStop}%
\bibitem [{\citenamefont {Fedichev}\ and\ \citenamefont
  {Fischer}(2003)}]{Fedichev2003}%
  \BibitemOpen
  \bibfield  {author} {\bibinfo {author} {\bibfnamefont {P.~O.}\ \bibnamefont
  {Fedichev}}\ and\ \bibinfo {author} {\bibfnamefont {U.~R.}\ \bibnamefont
  {Fischer}},\ }\href {\doibase 10.1103/PhysRevLett.91.240407} {\bibfield
  {journal} {\bibinfo  {journal} {Physical Review Letters}\ }\textbf {\bibinfo
  {volume} {91}},\ \bibinfo {pages} {240407} (\bibinfo {year}
  {2003})}\BibitemShut {NoStop}%
\end{thebibliography}
\end{document}